\documentclass{aa}  

\usepackage{graphicx}
\usepackage{txfonts}
\usepackage{lineno}
\usepackage{subfigure}
\usepackage{booktabs}
\usepackage{xcolor} % Abel: added on 21.05.2021
\usepackage{amsmath}
\usepackage{academicons}
\usepackage{hyperref}
\usepackage{orcidlink}
\usepackage{threeparttable}  % In preamble
\usepackage{float}
\usepackage{makecell}
\usepackage{tabularx,threeparttable,booktabs,makecell}
\newcommand{\orcid}[1]{\href{https://orcid.org/#1}{\textcolor[HTML]{A6CE39}{\aiOrcid}}}

\newcommand\kms{{\rm\,km\,s^{-1}}}

 \def\mso{\,\mathrm{M}_\odot}

 \def\simle{\mathrel{\hbox{\rlap{\hbox{\lower4pt\hbox{$\sim$}}}\hbox{$<$}}}}
 \def\simgr{\mathrel{\hbox{\rlap{\hbox{\lower4pt\hbox{$\sim$}}}\hbox{$>$}}}}
\renewcommand{\thefigure}{\arabic{figure}}

\begin{document}

   \title{
   Thermal-timescale accretion does not always yield critical rotation in mass gainers
   }

   \author{Chen Wang 
          \inst{1,2,3}
, Mike Y.M. Lau \inst{4,5}, Xiang-Dong Li \inst{1,2}, Norbert Langer \inst{6}, Selma E. de Mink \inst{3},
         Ruggero Valli \inst{3},  Stephen Justham \inst{3}, Xiao-Tian Xu \inst{1,2,7,6}, Jakub Klencki \inst{3},
          \and
          Taeho Ryu \inst{8,9,3}
          }

  \institute{School of Astronomy and Space Science, Nanjing University, Nanjing, 210023, People's Republic of China\\
              \email{chen\_wang@nju.edu.cn}
              \thanks{}
              \and
Key Laboratory of Modern Astronomy and Astrophysics, Nanjing University, Ministry of Education, Nanjing, 210023, People's Republic of China
\and
Max Planck Institute for Astrophysics, Karl-Schwarzschild-Strasse 1, 85748 Garching, Germany
\and
Zentrum f\"ur Astronomie der Universit\"at Heidelberg, Astronomisches Rechen-Institut, M\"onchhofstr. 12-14, 69120 Heidelberg
\and
Heidelberger Institut f\"ur Theoretische Studien, Schloss-Wolfsbrunnenweg 35, 69118 Heidelberg, Germany
\and
Argelander-Institut f\"ur Astronomie, Universit\"at Bonn, Auf dem H\"ugel 71, 53121 Bonn, Germany
\and
Tsung-Dao Lee Institute, Shanghai Jiao-Tong University, Shanghai, 201210, China
\and
JILA, University of Colorado and National Institute of Standards and Technology, 440 UCB, Boulder, 80308, CO, USA
\and
Department of Astrophysical and Planetary Sciences, 391 UCB, Boulder, 80309, CO, USA             
}

   \date{Received: November 13, 2025; Accepted: January 13, 2026}
\titlerunning{The initial spin distribution of B-type stars}
\authorrunning{Wang et al.}
% \abstract{}{}{}{}{} 
% 5 {} token are mandatory
 
  \abstract
   %context heading (optional)
   %{} leave it empty if necessary  
   {Binary evolution plays a central role in producing rapidly rotating stars. Previous studies have shown that mass gainers in binaries can reach critical rotation after accreting only modest amounts of material, particularly during thermal-timescale Case B mass transfer, where tidal spin-down is ineffective due to wide orbits. However, such rapid accretion often drives the mass gainer out of thermal equilibrium, and its subsequent spin evolution during thermal relaxation has not been analysed in depth. In this study, we construct a suite of accreting detailed single-star models with different accretion prescriptions, which inflate and spin up to critical rotation during the accretion. After the accretion has ended, the models relax thermally and deflate. We find that the ratio of surface to critical angular velocity decreases to subcritical values during thermal contraction, with the magnitude of this decrease correlating with the degree of thermal disequilibrium at the end of accretion. 
This reduction in fractional critical rotation is even stronger when internal angular momentum transport is inefficient.
Detailed binary models show the same trend, indicating that the results from our toy single-star models also apply to real binary evolution. 
Our results highlight that binary mass transfer does not always produce critically rotating stars, but instead may yield a wide range of spin rates depending on the mass transfer and accretion history. Our findings offer new insights into the rotational properties of mass gainers in binaries, stellar merger products, and newly formed massive stars following accretion.
}


   \maketitle
%
%-------------------------------------------------------------------
%===================================================================================================
\section{Introduction} \label{sec:intro}
%===================================================================================================
The VLT-FLAMES survey revealed a bimodal distribution of rotational velocities for B-type stars in the Large Magellanic Cloud (LMC), with $v\sin i$ peaking at $\leq 100\,\kms$ and $\sim 250\,\kms$, and a long tail extending to $\sim 450\,\kms$ \citep{Dufton2013}. Similar distributions have been reported for Galactic B- and A-type stars \citep{Huang2010,Zorec2012}. The fast-rotating tail is widely interpreted as a consequence of binary interaction.

Among the rapid rotators, Be stars receive particular attention. These are B-type stars showing emission lines, typically attributed to circumstellar decretion disks formed due to near-critical rotation (see \citealt{2013A&ARv..21...69R} for a review). Compelling evidence for a binary origin of Be stars include the detection of Be X-ray binaries \citep{2005A&AT...24..151R,2006A&A...455.1165L}, Be+sdO systems \citep{2008ApJ...686.1280P,2017ApJ...843...60W,2021AJ....161..248W}, the observed lack of Be stars with main-sequence companions \citep{2020A&A...641A..42B}, and the presence of faint Be stars located below the main-sequence turnoff in young open star clusters \citep{Milone2018,2020ApJ...888L..12W,2023A&A...670A..43W,2020A&A...633A.165H,2021A&A...653A.144H}.

While binary evolution is widely recognized as a key channel for producing rapidly rotating stars \citep{1981A&A...102...17P,1991A&A...241..419P,1997A&A...322..116V,2005A&A...435..247P,2007A&A...465L..29C,2013ApJ...764..166D}, the spin evolution of accretors after mass transfer remains insufficiently explored. Detailed binary models that account for both mass and angular momentum transfer typically predict that the accretor reaches critical rotation after accreting only a small fraction of its own mass, usually less than $\sim$5–10\% \citep{2020ApJ...888L..12W,2024arXiv241102376A,2025arXiv250323876X}. To prevent exceeding critical rotation, the accretion rate is then limited (hereafter rotation-limited models), leading to very low accretion efficiencies. In such models, the accretor remains close to critical rotation for the rest of its main-sequence lifetime unless stellar winds are sufficiently strong to spin it down.

However, recent studies suggest that higher accretion efficiencies than those predicted by rotation-limited models are required to explain the observed properties of Be+sdO binaries, Be X-ray binaries and the observed efficiency of forming Oe stars \citep{2007ASPC..367..387P,2014ApJ...796...37S,2018A&A...615A..30S,2020MNRAS.498.4705V,2025arXiv250514780L,2025arXiv250323878S,2025arXiv250323876X}. Studies of accreting stars that neglect stellar rotation, thereby permitting higher accretion efficiencies, show that accretors can become strongly inflated due to significant departures from thermal equilibrium (TE) during rapid mass transfer \citep{1976ApJ...206..509U,1977PASJ...29..249N,2007A&A...467.1181D,2024A&A...691A.174S,2024ApJ...966L...7L}. Such stars subsequently undergo thermal relaxation (i.e., contraction) either after the donor detaches or once the accretor grows sufficiently in mass and luminosity that its thermal timescale becomes shorter than the accretion timescale.
Yet, because these models neglect rotation, while rotation-limited models limit the extent of departure from TE, the spin evolution of thermally disequilibrated stars during the post-accretion relaxation phase has not been systematically investigated.

There are indications that spin-down can occur during the thermal relaxation phase. For example, the accretor in the detailed binary model of \citet{2021ApJ...923..277R} spins down from critical rotation to about 80\% of the critical angular velocity shortly after thermal-timescale accretion ceases, which they attributed to strong stellar winds. However, when we recomputed a comparable model at lower metallicity and lower mass  (see Appendix\,\ref{app_sec:E}), where wind mass loss is negligible, we still found spin-down during the thermal relaxation phase, indicating that winds are not the primary cause. We emphasize that throughout this work, we define “spin-down” as a decrease in the ratio of surface angular velocity to the critical angular velocity $\omega_{\mathrm{crit}}=\sqrt{(1-\Gamma)GM/R^3}$, where $M$, $R$ and $\Gamma$ denote the stellar mass, radius, and Eddington factor, respectively, rather than a reduction in the absolute surface angular velocity nor a reduction of the total angular momentum. When studying the role of rotation in stellar evolution, particularly in the formation of Be stars, this ratio is the more physically relevant quantity, as it determines how easily material can be ejected. 
%Addition clues come from the study of main-sequence merger products by \citet{2019Natur.574..211S}, who found that the merger product initially rotates at the critical velocity but spins down to just $\sim$10\% of critical rotation after regaining TE.

In this study, we investigate the spin evolution of stars during the thermal relaxation phase that follows thermal-timescale mass transfer. We do not consider the nuclear-timescale Case A mass transfer, because in that case the accretor remains in thermal equilibrium throughout accretion, and tidal interactions in short-period binaries can efficiently modify the stellar spin. 
We construct a suite of toy single-star models in which mass accretion drives the stars to different levels of thermal disequilibrium, and then follow their subsequent spin evolution. These toy models allow us to identify the underlying processes driving stellar spin evolution during thermal relaxation. We then compare these results with detailed binary models.

This paper is organized as follows. In Section\,\ref{sec:method}, we describe the model sets and physical assumptions. Section\,\ref{sec:Results} presents the spin evolution of stars during thermal relaxation. In Section\,\ref{sec:discussion}, we discuss the broader implications of our results on binary evolution, binary mergers and star formation. Finally, we summarize our main conclusions in Section\,\ref{sec:conclusion}.

%In this study, we define spin-down as a decrease in the ratio \omega / \omega_{\mathrm{crit}}, not necessarily a decrease in the absolute angular velocity. 

\begin{table*}[t]
\caption{Summary of the initial parameters and physics assumptions of our models.}
\centering
\begin{minipage}{\textwidth}
\centering
\begin{threeparttable}
\footnotesize
% Make the table exactly \textwidth so notes match it
\begin{tabularx}{\textwidth}{l@{\hspace{-6pt}} c@{\hspace{-5pt}} c@{\hspace{-6pt}} c@{\hspace{-6pt}} c@{\hspace{-3pt}} c@{\hspace{6pt}}}
\toprule
\midrule
Single-star model  & $m_\mathrm{i}\,(M_\odot)$ &
$X_\mathrm{H}$ ($t=t_\mathrm{a}$) &
$\log \dot{M}_\mathrm{acc,i}\,(M_\odot/\mathrm{yr})$ &
$\dot{M}_\mathrm{acc}$  & accrertion termination condition  \\
\midrule
Set I    & 4–20, $\Delta M=2$ & 0.5 & -4.3 to -2.2, in steps of 0.1 & constant & critical rotation \\
Set II   & 8  & 0.5 & -3.0 & rotation-limited & $\log \dot{M}_\mathrm{acc}$ drops below certain thresholds \\
Set III  & 8  & 0.3 & -3.0 & rotation-limited &$\log \dot{M}_\mathrm{acc}$ drops below certain thresholds \\
Set IV   & 12 & 0.5 & -3.3 and -3.0 & not rotation-limited & mass reaches $20\,\mso$ \\
\midrule \midrule
Binary model  & initial binary parameters & $\alpha_\mathrm{SC}$ & $j_\mathrm{acc}$ & rotation-limited $\dot{M}_\mathrm{acc}$? & Tayler–Spruit dynamo\\
\midrule
Model I   & $(20+8)\,\mso$, 150 d & 1   & $j_\mathrm{kep}$                     & yes & yes \\
Model II  & $(20+8)\,\mso$, 150 d & 1   & $j_\mathrm{accretor}$                & yes & yes \\
Model III & $(20+8)\,\mso$, 150 d & 100 & $j_\mathrm{kep}$                     & yes & yes \\
Model IV  & $(20+8)\,\mso$, 150 d & 1   & $j_\mathrm{kep}$ when $\omega<\omega_\mathrm{crit}$ & no  & yes \\
  &  &   & $0.1\,j_\mathrm{kep}$ when $\omega=\omega_\mathrm{crit}$ &  & \\
Model V   & $(20+8)\,\mso$, 150 d & 1   & $j_\mathrm{kep}$                     & yes & no  \\
\bottomrule
\end{tabularx}

% Notes now share the same \textwidth via the minipage/threeparttable measurement
\begin{tablenotes}[para,flushleft]
\footnotesize
\item $m_\mathrm{i}$: initial mass; $X_\mathrm{H}$: central hydrogen mass fraction; $t_\mathrm{a}$: time when accretion begins; $\dot{M}_\mathrm{acc,i}$: initial mass accretion rate;
$\alpha_\mathrm{SC}$: semiconvective mixing parameter; $j_\mathrm{acc}$: specific angular momentum accreted by the star;
$j_\mathrm{kep}$: Keplerian specific angular momentum at the accretor's surface; $j_\mathrm{accretor}$: specific angular momentum at the surface of the accretor. $\omega$ and $\omega_\mathrm{crit}$: surface angular velocity and critical angular velocity, respectively.
\end{tablenotes}

\end{threeparttable}
\end{minipage}
\end{table*}

%===================================================================================================
\section{Stellar models and physics assumptions}\label{sec:method}
%===================================================================================================
We use the 1D stellar evolution code MESA (version 8845, \citealt{Paxton2011,Paxton2013,Paxton2015}) to compute both single and binary evolution models. We adopt low metallicity representative of SMC stars to minimize the influence of stellar winds on spin evolution. The physics assumptions follow those in \citet{2020ApJ...888L..12W}. All single-star models are initialized with a rotational velocity of $50\,\kms$, while the binary models are assumed to be tidally synchronized with their orbits initially. The initial parameters and accretion prescriptions for each model are summarized in Table\,1.
\subsection{Single-star models}
We construct four sets of single-star models that undergo mass accretion under different prescriptions. We firstly investigate the effect of constant mass accretion rate with Set I, which consists of stellar models with initial masses ranging from 4 to 20$\mso$ in steps of 2$\mso$. These models are evolved from the zero-age main-sequence until their central hydrogen mass fraction drops to 0.5, at which point they begin accreting mass at constant rates with $\log_{10} (\dot{M}/\mathrm{M_\odot\,yr^{-1}})$ ranging from $-4.3$ to $-2.2$ in steps of 0.1. We terminate the accretion once the star reaches critical rotation. The models are then evolved further until the central hydrogen mass fraction drops to 0.48, shortly after the star regains TE. The accreted material is assumed to carry the same specific angular momentum, specific entropy, and chemical composition as the stellar surface.

Models in Set II explore a more realistic scenario in which accretion does not stop abruptly at critical rotation. Previous detailed binary evolution models with rotation-limited accretion have shown that after a star first reaches critical rotation, it can continue accreting at a reduced rate and remain near the critical limit. This is possible because stellar contraction gradually raises the critical rotation threshold (\citealt{2020ApJ...888L..12W}, see also Appendix\,\ref{app_sec:E}).
In our Set II models, an 8$\mso$ star starts accreting when its central hydrogen mass fraction drops to 0.5, at a rate of $1.0\times10^{-3}\,M_\odot\,\mathrm{yr^{-1}}$, until it first reaches critical rotation. Thereafter, the accretion rate is implicitly calculated to keep the star just below the critical limit. Accretion is manually terminated once the rate falls below a prescribed threshold, with $\log (\dot{M} / \mathrm{M_\odot\,yr^{-1}})$ ranging from $-4.75$ to $-3.05$ in steps of 0.25. Thermal contraction begins soon after the star first reaches critical rotation and the accretion rate starts to decline (i.e. before accretion fully ceases). Therefore, terminating accretion later, at a lower $\dot{M}$ value means the accretor has contracted further and is therefore closer to TE at the end of accretion, and vice versa. We terminate the calculation when the central hydrogen mass fraction reaches 0.48.

Set III is identical to Set II in terms of physics and accretion prescriptions, but explores the role of the accretor's evolutionary stage by shifting the timing of accretion to later in the main sequence. Accretion begins when the central hydrogen mass fraction reaches 0.3 and is manually terminated at the same accretion-rate thresholds as in Set II. Each model is evolved until the central hydrogen mass fraction drops to 0.28.

Finally, motivated by recent observations that support for higher accretion efficiency, as discussed in Section\,\ref{sec:intro}, we construct toy models in which accretion efficiency can be maintained at a relatively high value by reducing the specific angular momentum of the accreted material by a factor of ten once the star first hits critical rotation. Two models are computed in Set IV: a 12\,$\mso$ star accreting at initial rates of $1.0 \times 10^{-3}\,M_\odot\,\mathrm{yr}^{-1}$ and $5.0 \times 10^{-4}\,M_\odot\,\mathrm{yr}^{-1}$, starting at central hydrogen mass fractions of 0.5, with accretion terminated when the star reaches 20\,$\mso$. While this reduction factor has no physical basis, it allows the star to maintain near-critical rotation while accreting at a relatively high rate. In the absence of detailed stellar models that treat this process self-consistently, our toy models serve as a first attempt to explore spin evolution when accretion continues beyond the critical threshold.

\subsection{Binary models}
We compute five detailed binary evolutionary models using the MESA code, with different physical assumptions but the same initial binary configuration: a primary mass of $20\,\mso$, a mass ratio of 0.4, and an orbital period of 150 days.  

The first three correspond to the physical assumptions adopted in binary models discussed in \citet{2020ApJ...888L..12W}, \citet{2021ApJ...923..277R}, and \citet{2022A&A...662A..56K}, respectively. 
%While not exact replicas, we recompute three detailed binary models with different physical assumptions, chosen to represent the setups in these studies. 
In Models\,I and III, the accreted material is assumed to carry the Keplerian specific angular momentum at the stellar surface. In contrast, Model\,II follows \citet{2021ApJ...923..277R}, assuming that the accreted material has the same specific angular momentum as the stellar surface (as in our single-star models), which is lower than in Models\,I and III. Models\,I and II adopt a semiconvective mixing efficiency parameter of 1, while Model\,III uses a higher value of 100. This higher efficiency leads to a partially stripped star and a longer Case\,B mass-transfer phase, representing the model of \citet{2022A&A...662A..56K}.  

Model\,IV is a toy model in which we artificially reduce the specific angular momentum of the accreted material by a factor of 10 when the accretor first reaches critical rotation, analogous to the setup in our single-star model Set\,IV. All other assumptions are the same as in binary Model\,I. This model provides a test case where mass accretion can continue even after the accretor reaches critical rotation.  

In Model\,V, we switch off the Tayler–Spruit dynamo, given the ongoing debate about its validity in massive stars. All other physics assumptions are the same as in Model\,I. This setup allows us to investigate the spin evolution of accretors when internal angular momentum transport in the radiative regions is very inefficient.  

%===================================================================================================
%===================================================================================================
\section{Stellar spin-down during thermal relaxation}\label{sec:Results}

\begin{figure*}
\includegraphics[width=\linewidth]{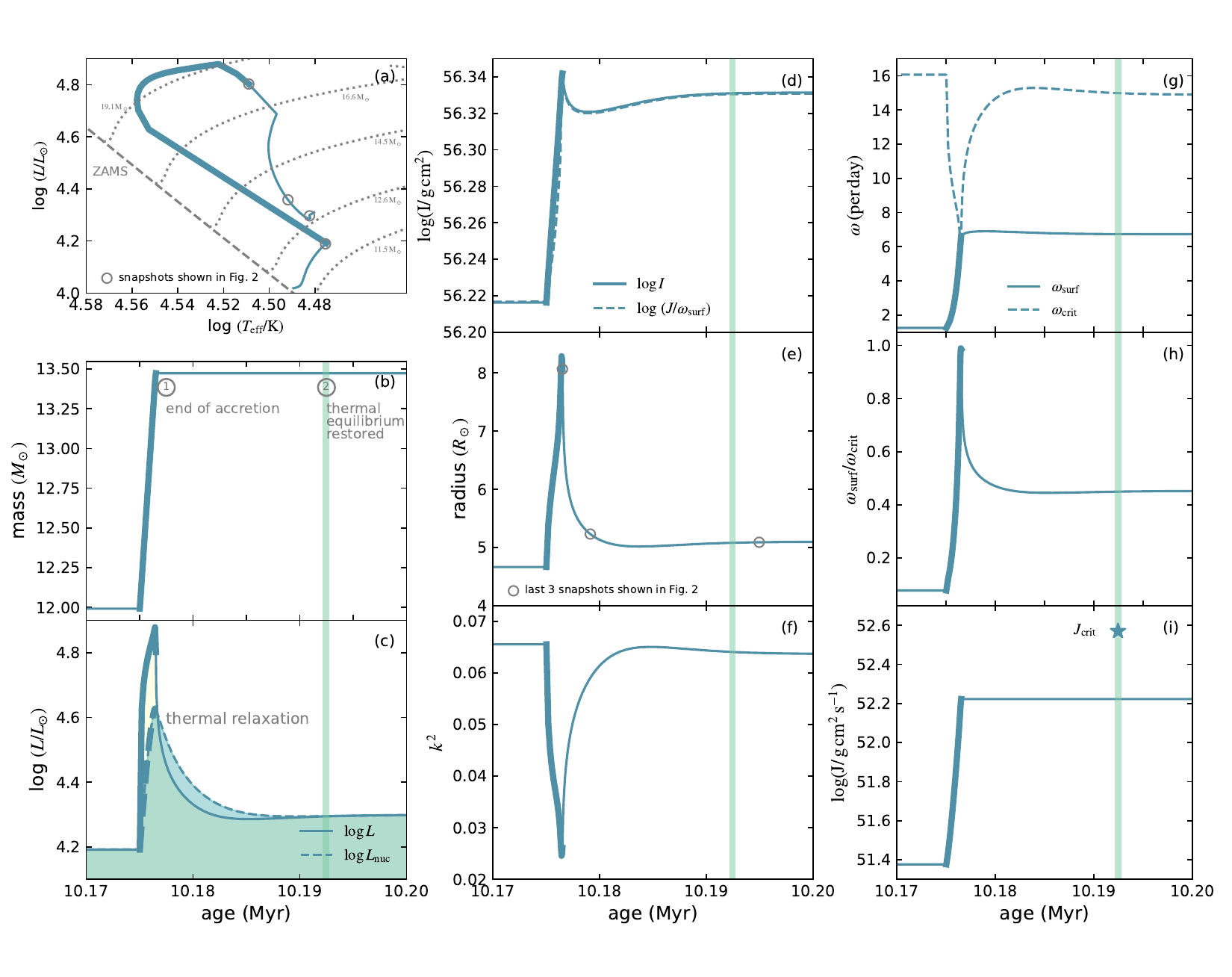}
\caption{Evolution of a 12$\mso$ stellar model from Set I 
with an accretion rate of $1.0\times 10^{-3}\,\mathrm{M_\odot \,yr^{-1}}$. Accretion begins when the stellar central hydrogen mass fraction drops to 0.5 and ends when the star reaches critical rotation. Panel (a) shows the model’s evolution in the Hertzsprung–Russell diagram. The grey dashed line marks the positions of non-rotating zero-age main-sequence stars, and grey dotted lines show evolutionary tracks of non-rotating single-star models from \cite{2019A&A...625A.132S}. Open circles denote the four snapshots for which 2D visualizations of the stellar radius and enclosed mass are presented in Fig.\,\ref{fig:12Msun_pie}. Panels (b)–(i) display the time evolution of key physical quantities throughout mass accretion and subsequent thermal relaxation. Thick segments correspond to the accretion phase, and the vertical green line marks the time when thermal equilibrium (TE) is restored. Here we define the restoration of TE as the time when the difference between the nuclear energy generation rate and the stellar luminosity falls below 0.2\%. 
(b) stellar mass; Two key evolutionary times-the end of accretion (1) and the restoration of TE (2) are indicated. (c) stellar luminosity (solid line) and luminosity from nuclear burning (dashed line); To facilitate comparison between the two quantities, the regions beneath the curves are shaded in different colors. (d) moment of inertia $I$ (solid) and the logarithm of the ratio of the total angular momentum $J$ to the surface angular velocity $\omega$ (dashed). (e) stellar radius; Open circles correspond to the last three snapshots shown in Fig.\,\ref{fig:12Msun_pie}. (f) gyration constant $k^2$. (g) stellar surface angular velocity (solid) and critical angular velocity (dashed). (h) ratio of surface angular velocity to the critical value.  (i) total angular momentum (solid) and the angular momentum required to maintain critical rotation once thermal equilibrium is restored assuming solid-body rotation (blue asterisk).}
\label{fig:fig1}
\end{figure*}

In this section, we present the spin evolution of stars during accretion and the subsequent thermal relaxation predicted by our stellar models, along with the underlying physical explanations.
Figure\,\ref{fig:fig1} shows the evolution of a 12$\mso$ stellar model from Set I with a constant accretion rate of $1.0\times 10^{-3}\,\mathrm{M_\odot \,yr^{-1}}$. Here we briefly summarize the key results. More details and other models from Set I are provided in Appendix\,\ref{app_sec:B}.

During the accretion phase, the total luminosity exceeds the nuclear luminosity (Panel c), indicating that the star departs from TE, which causes it to expand (Panel e).
Meanwhile, the surface angular velocity $\omega$ increases, while the surface critical angular velocity $\omega_\mathrm{crit}$ decreases due to expansion (Panel g). Together, these effects drive the star toward critical rotation (Panel h). Once critical rotation is reached, accretion is terminated and the star enters the thermal relaxation phase.

During thermal relaxation, the star contracts to restore TE (Panel e). 
The surface angular velocity remains nearly constant, while the critical angular velocity increases as the stellar radius decreases (Panel g), resulting in a decline of the ratio $\omega/\omega_\mathrm{crit}$ (Panel h). Thus, during thermal relaxation, it is the fractional critical rotation that decreases, rather than the absolute angular velocity. However, since the stellar radius shrinks, the absolute linear velocity at the surface decreases. As mentioned in Section\,\ref{sec:intro}, we refer to this decline in fractional critical rotation as spin-down in this study. The spin-down discussed here is relative to the moment when accretion ceases, rather than to the star's initial rotation, since angular momentum accretion inevitably spins the star up compared to its pre-accretion state.

\begin{figure*}
\includegraphics[width=\linewidth]{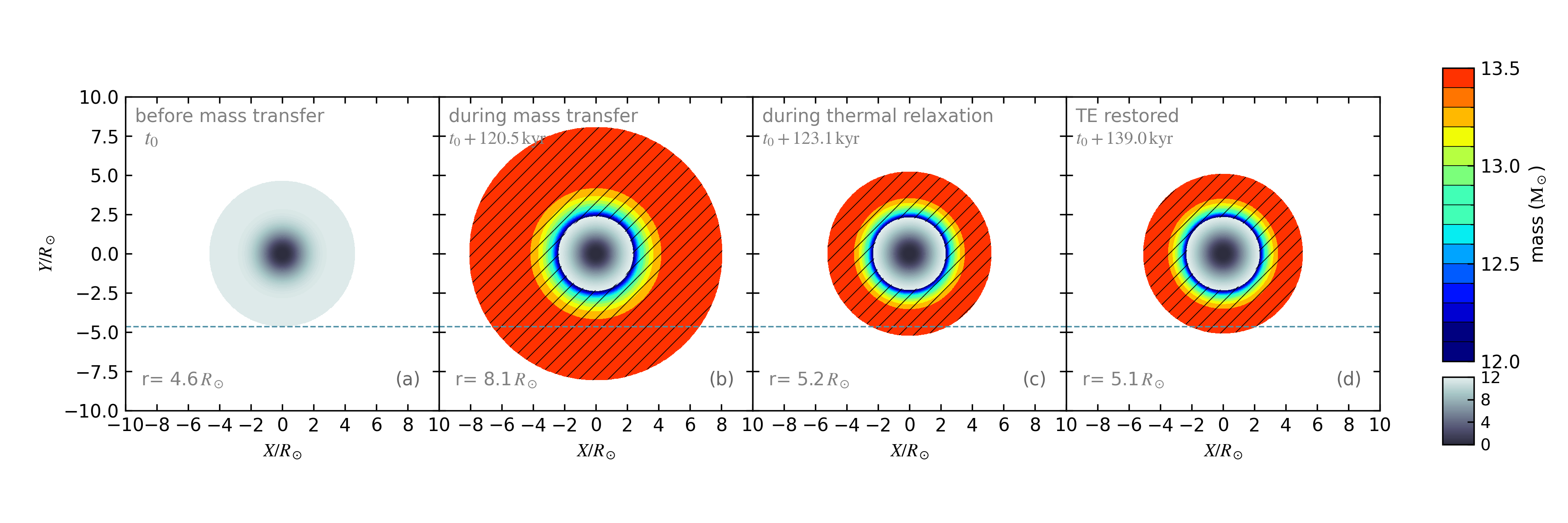}
\caption{2D visualization of radius and enclosed mass at four snapshots for a 12$\mso$ star from single-star model Set I, with an initial accretion rate of $1.0\times 10^{-3}\,\mathrm{M_\odot\,yr^{-1}}$. The snapshots span evolution phases before mass transfer (MT), during MT, during thermal relaxation (TR), and after restoring thermal equilibrium (TE). The reference time is $t_0=10.06\,$Myr. We ignore deviations from spherical symmetry due to critical rotation and use the averaged radius in the X-Y plane. The color scale denotes enclosed mass, with two distinct colormaps applied below and above $12\,\mso$ to differentiate the original stellar mass from the accreted mass.
Hatched regions in Panels (b)–(d) highlight the accreted layers. The thin horizontal dashed line marks the stellar radius at $t_0$, and the current stellar radius is labeled in each panel.}
\label{fig:12Msun_pie}
\end{figure*}

Two factors explain why $\omega$ remains almost constant during thermal relaxation. First, the stellar moment of inertia $I$ changes only marginally during this phase (approximately 2\%, see Panel d). 
%The moment of inertia is computed as
%$$I = \int_{V}(r\sin\theta)^2\mathrm{d}m,$$
%where $\mathrm{d}m = \rho(r)\,r^2\sin\theta\,\mathrm{d}r\,\mathrm{d}\theta\,\mathrm{d}\phi$ and $\rho(r)$ is the density at radius coordinate $r$.
%Integrating over the spherical volume yields
%$$I = \int_0^{2\pi}\!\!\int_0^\pi\!\!\int_0^R \rho(r)\,r^4\sin^3\theta\,\mathrm{d}r\,\mathrm{d}\theta\,\mathrm{d}\phi=\frac{8\pi}{3} \int_{0}^{R} \rho(r) \, r^{4} \, \mathrm{d}r,$$
%where $R$ is the stellar radius.
We find that the stellar structure evolves non-homologously during accretion and thermal relaxation. The envelope undergoes dramatic radial expansion and contraction, whereas the core remains largely unchanged. This can be seen clearly in Fig.\,\ref{fig:12Msun_pie}. 
This behavior arises because the radiative envelope is inefficient at transporting energy, so the excess energy deposited by accretion is trapped within it, driving its expansion. In contrast, the convective core transports energy efficiently, allowing it to maintain near-equilibrium without significant changes (see Fig.\,\ref{fig:12Msun_profile} for details).

Most of the mass and angular momentum reside in the dense central regions, whereas the inflated envelope is very tenuous and contains negligible mass (the outer 60\% of the stellar radius contains only $\sim$2.7\% of the total mass in this model), angular momentum and moment of inertia (see Fig.\,\ref{fig:12Msun_pie} and Fig.\,\ref{fig:12Msun_profile}). Thus the thermal contraction, which mainly affects the inflated envelope, leaves the total moment of inertia largely unchanged (with a variation of only about 2.5\%). 

The second reason $\omega$ remains almost constant is that at this moderate accretion rate ($1.0\times 10^{-3}\,\mathrm{M_\odot \,yr^{-1}}$, corresponding to an accretion timescale of about 14\% of the thermal timescale), angular momentum transport processes are operating fast enough to allow the star to rotate nearly rigidly during accretion. This can be seen from Panel d in Fig.\,\ref{fig:fig1}, where the moment of inertia approximately equals the ratio of the total angular momentum to the surface angular velocity, and from the nearly flat $\omega$ profiles shown in Panel g of Fig.\,\ref{fig:12Msun_profile}.
At the end of accretion, the core and envelope rotate at similar angular velocities, close to the mean velocity, so no significant angular momentum redistribution is required. With angular momentum conserved and moment of inertia remaining nearly constant, the surface angular velocities change only marginally.

The gyration constant $k^2$, defined through $I = k^2MR^2$, where $M$ and $R$ are stellar mass and radius, respectively, measures the overall central concentration of the star (Panel f in Fig.\,\ref{fig:fig1}). During accretion, $k^2$ decreases, reflecting a more centrally concentrated configuration, whereas during thermal relaxation, $k^2$ increases, indicating a less centrally concentrated structure compared to the accretion phase. This variation reflects the change in stellar radius in the absence of substantial change to the moment of inertia.

The above description applies to cases with relatively low accretion rates, for which rigid rotation is maintained. At very high accretion rates, e.g., $4.0\times10^{-3}\,M_\odot\,\mathrm{yr^{-1}}$ for a $12\,M_\odot$ star (corresponding to an accretion timescale of about 3\% of the thermal timescale), even the inclusion of the Taylor–Spruit dynamo cannot enforce rigid rotation during accretion (see Panel g in Fig.\,\ref{fig:12Msun_profile_2.4}). In such cases, the stellar envelope rotates faster than the core at the end of accretion. During thermal relaxation, although the total moment of inertia remains nearly constant, angular momentum is transported inward, reducing the surface angular velocity. Combined with the simultaneous increase of $\omega_\mathrm{crit}$, this results in a stronger spin-down than in the rigid rotation case.

The evolution of $\omega$ and $\omega_\mathrm{crit}$ show the spin-down of post-accretion stars during thermal relaxation.
The more fundamental reason is that envelope inflation limits the amount of angular momentum that can be accreted, such that it is sufficient to sustain critical rotation only while the star is in its expanded configuration, but not once it contracts back into TE (see Panel i in Fig.\,\ref{fig:fig1}).

\begin{figure*}
\includegraphics[width=\linewidth]{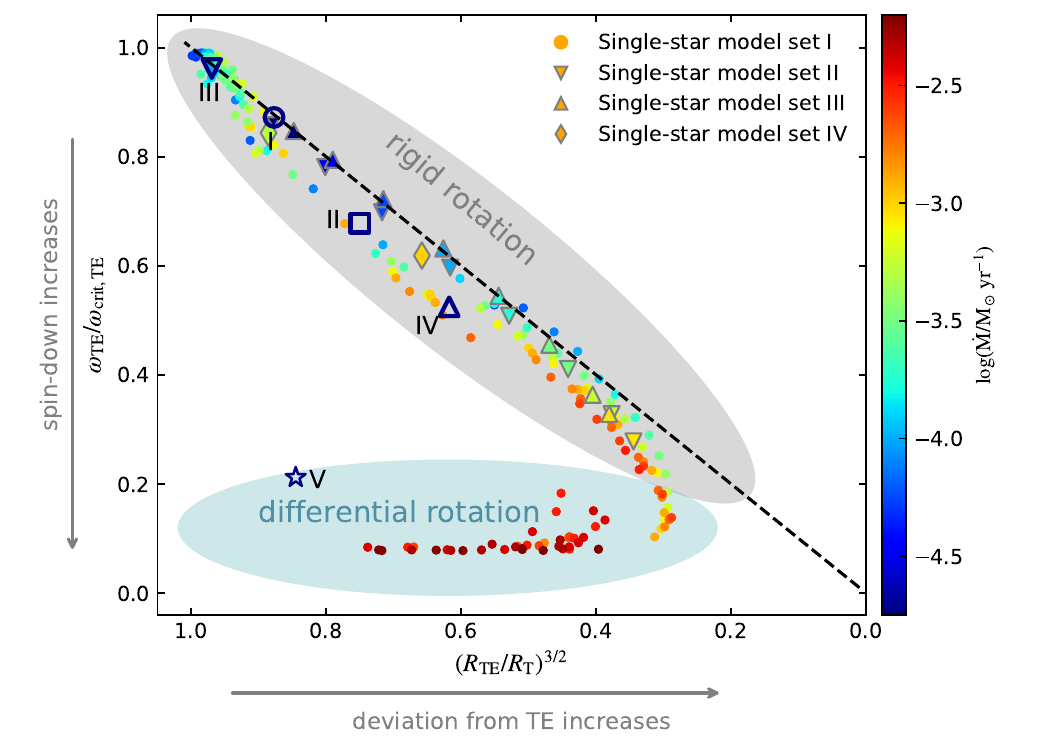}
\caption{Relationship between stellar spin-down during thermal relaxation and the degree of thermal disequilibrium at the time mass accretion ceases, for all models computed in this study. The spin-down is characterised by $\omega_\mathrm{TE}/\omega_\mathrm{crit,TE}$, where $\omega$ and $\omega_\mathrm{crit}$ denote the surface and critical angular velocities, respectively, and the subscript ``TE" refers to the time when thermal equilibrium (TE) is restored. The deviation from TE is quantified by $(R_\mathrm{TE}/R_\mathrm{T})^{3/2}$, where $R_\mathrm{T}$ and $R_\mathrm{TE}$ are the stellar radii at the termination of accretion and at the restoration of TE, respectively. The exponent $3/2$ arises from the dependence of the critical angular velocity on stellar radius (see Appendix\,\ref{app_sec:A} for details).
Filled circles without grey edges represent single-star models from Set I, while filled markers with grey edges correspond to single-star models from other sets, with different markers denoting different sets, as indicated in the legend.
Open markers represent binary models, with model number labeled beside. 
The color of each marker (or its edge color for binaries) indicates the mass accretion rate at the termination of accretion.
The black dashed line indicates a linear relation with a slope of -1. The grey and cyan shaded regions indicate models that rotate rigidly and differentially, respectively, at the termination of accretion. }
\label{fig:v_k}
\end{figure*}

Since an increase in the critical rotation velocity leads to a decrease in the fractional critical rotation rate, we expect that a larger deviation from TE at the end of mass accretion will result in a greater change in stellar radius and critical velocity during thermal relaxation, and consequently, a stronger spin-down. In Fig.\,\ref{fig:v_k}, we present the relationship between $\omega_\mathrm{TE}/\omega_{\mathrm{crit,TE}}$ and $(R_\mathrm{TE}/R_\mathrm{T})^{3/2}$ (see Appendix\,\ref{app_sec:A} for more details) for all single and binary models computed in this study. Here the subscripts T and TE correspond to the time when mass accretion terminates and when TE is restored, respectively. The ratio $R_\mathrm{TE}/R_\mathrm{T}$ measures the relative change in stellar radius between these two stages, with smaller values corresponding to a larger departure from TE. The exponent 3/2 arises from the dependence of the critical angular velocity on stellar radius, $\omega_\mathrm{crit} \propto R^{-3/2}$.

Two groups can be distinguished. The first group contains models closely follow the linear relation $\omega_\mathrm{TE}/\omega_{\mathrm{crit,TE}}\propto(R_\mathrm{TE}/R_\mathrm{T})^{3/2}$. These models accrete at relatively low rates and near-rigid rotation is maintained. There are models with relatively high accretion rates that follow this linear relation, for example, the higher-mass stars in single-star model Set I (see Fig.\,\ref{fig:stop_at_crit_rot} for more details). This is because higher-mass stars can maintain near-rigid rotation at higher mass accretion rates. 
%For stellar models in this group, the surface angular velocity changes little during thermal contraction, and the spin-down is primarily caused by the increase in $\omega_\mathrm{crit}$. This naturally explains the linear relation: larger deviations from TE imply greater contraction during thermal relaxation and therefore a larger increase in $\omega_\mathrm{crit}$, leading to a stronger spin-down.

The second group consists of models with higher accretion rates, in which rigid rotation breaks down and $\omega_\mathrm{TE}/\omega_\mathrm{T}$ significantly deviates from unity. These models deviate from the above-mentioned relation (see Appendix\,\ref{app_sec:A} for the expression). During thermal relaxation, their surface angular velocity decreases, and the combined effects of radius contraction and reduced surface rotation lead the star to rotate at a smaller fraction of the critical velocity for a given departure from TE.
A notable outlier is binary model V, in which angular momentum transport in radiative zones is inefficient due to the absence of the Tayler–Spruit dynamo.
In this model, only the outermost surface layers are spun up, and the star accretes very little mass before reaching critical rotation, producing a minimal deviation from TE.
The subsequent spin-down is dominated by internal angular momentum redistribution from the surface to the core (see Appendix\,\ref{app_sec:G}).

%An analytical explanation is provided in Appendix,\ref{app_sec:A}, where we show that this relation arises naturally from conservation of angular momentum.

%===================================================================================================
\section{Discussion}\label{sec:discussion}
\subsection{Implications for binary evolution}\label{sec:discussion1}
\subsubsection{Post-accretion stars may exhibit diverse rotation rates}
From our toy single-star models, we find that the key parameter determining the rotation rate of an accretor when it regains TE is the extent of its deviation from TE at the end of accretion: larger deviations lead to stronger post-accretion spin-down. Real binary evolution, however, is far more complex. The mass accretion history in binaries is governed by multiple factors, including the mass-transfer behavior, how easily the accretor reaches critical rotation, and what happens once that limit is reached.
In general, mass transfer increases rapidly and then gradually declines during the thermal-timescale mass-transfer phase (see the orange lines in the middle panels of Fig.\,\ref{app_fig:binary_main}). The accretion behavior, namely whether it follows or departs from the evolution of mass-transfer rate (i.e., the mass accretion efficiency), depends sensitively on uncertain aspects of binary evolution physics, particularly the specific angular momentum of the accreted material and the treatment of accretion once the accretor reaches critical rotation. These uncertainties ultimately determine how far the accretor departs from TE at the end of mass accretion. As a result, it remains fundamentally difficult to predict the final rotation rate of an accretor in any given binary system.

In our rotation-limited binary models (I, II, and III), the accretion rate follows the mass-transfer rate until the accretor first reaches critical rotation (see the middle panels in Fig.\,\ref{app_fig:binary_main}). Once this point is reached, the accretion rate begins to decline and no longer drives the star further from TE. Thus, the maximum deviation from TE during accretion is set by the maximum accretion rate the star can reach, which in turn depends sensitively on the assumed specific angular momentum of the accreted material. A lower specific angular momentum (as in binary Model II) leads to a higher peak accretion rate, since the star must accrete more to reach critical rotation, and consequently a larger deviation from TE compared to models with higher accreted specific angular momentum. Although it is likely unrealistic to assume that the accreted material has the specific angular momentum of the surface of the accretor, this remains the default assumption in the MESA code (i.e., do\_j\_accretion$=$.false.). 
 
During the accretion phase with a reduced rate, the star contracts while remaining at critical rotation. The rate at which accretion declines then plays a crucial role in determining how far the star departs from TE at the end of accretion. Assuming lower specific angular momentum for the accreted material leads to a slower decline in the accretion rate, keeping the star farther from TE by the end of mass transfer, and thus producing stronger spin-down afterward.

In closer binaries (for example, systems with periods shorter than approximately 15 days and the same component masses as our five binary models), mass accretion is expected to proceed via ballistic accretion \citep{1975ApJ...198..383L,1976ApJ...206..509U}, in which the specific angular momentum of the accreted material differs from that assumed in the models presented here. Nevertheless, our binary models I and II, which adopt different prescriptions for the specific angular momentum of the accreted material, already demonstrate that as long as the accretor is spun up by accretion and driven out of thermal equilibrium, its subsequent spin evolution during the thermal relaxation phase follows the same trends identified in this study.
In addition, for such close binaries, the post-mass-transfer orbital separation may be sufficiently small that tidal interactions can further modify the spin of the accretor.
A comprehensive exploration of more realistic angular momentum prescriptions is beyond the scope of this work. Instead, our goal here is to demonstrate how different assumptions about angular momentum transfer shape the accretion history and the spin evolution of the accretor.

In our detailed binary models, the end of accretion is controlled by the termination of mass transfer, which itself depends on the donor’s structural response to mass loss.
In particular, a higher semi-convective mixing efficiency (binary Model III) results in a longer mass-transfer phase, because the donor is only partially stripped, and the slow expansion of the remaining envelope continues to drive mass transfer (see \citealt{2022A&A...662A..56K} for details). A longer-lasting mass-transfer episode allows the accretor more time to contract and thus approach TE more closely by the end of accretion, resulting in a weaker spin-down.
Alternatively, accretion may stop abruptly before mass transfer ends, for example, via L2 overflow. Even though L2 overflow does not occur in our binary models, studies suggest that in systems with rapidly rotating accretors, such as Be stars, the L2 geometry deviates from that of non-rotating binaries \citep{2007ApJ...660.1624S}.

The above mentioned discussion applies to rotation-limited accretion. Under this assumption, the accretor in Model II already spins down significantly, reaching about 68\% of its critical rotation after regaining TE.
However, recent studies indicate that such models cannot reproduce many observed systems, such as Be+sdO and Be X-ray binaries, which require higher accretion efficiencies \citep{2007ASPC..367..387P,2014ApJ...796...37S,2018A&A...615A..30S,2025arXiv250514780L}. Higher accretion efficiencies may be achieved if accretion continues after the star reaches critical rotation through the outward transport of angular momentum via a disk \citep{1991ApJ...370..604P,1991ApJ...370..597P,1991MNRAS.253...55C}. In this scenario, the accretion rate may reach higher values and remain high for a longer period compared to rotation-limited models. Both effects cause a stronger spin-down of the accretor during thermal relaxation. Due to the lack of detailed binary models that include this process self-consistently, we constructed a toy binary model (binary Model IV) that artificially reduce the specific angular momentum of the accreted material after the star reaches critical rotation. As expected, the accretor in this model experiences a stronger post-accretion spin-down than in rotation-limited cases, reaching only about 52\% of the critical velocity, which is below the threshold required to appear as a Be star.

The spin reduction of accretors in binaries is less dramatic than in some of our single-star experiments because mass transfer rate declines gradually rather than stopping abruptly. Even in the most extreme case of fully conservative mass transfer (where the accretion rate equals the mass-transfer rate throughout the interaction), the transfer rate eventually drops enough for the mass-transfer timescale to become comparable to the accretor’s thermal timescale (see Fig.\,\ref{app_fig:binary_timescale}). As a result, the accretor begins to contract before mass transfer completely ceases. Whether the star spins down to a sub-critical state then depends on whether it can regain TE during the remaining mass-transfer phase.
In our binary Model II, the remaining accretion time after the two timescales become comparable is only less than a thousand years (see Fig.\,\ref{app_fig:binary_timescale}), which is much shorter than the thermal timescale of several tens of thousands of years.
This indicates that the star cannot fully regain TE by the end of mass transfer, and therefore it will rotate sub-critically after thermal readjustment, assuming conservative mass transfer.

Overall, our results indicate that binary evolution can produce a broad range of post-accretion rotation rates, including values below the threshold required for a Be-star appearance, rather than driving all accretors to critical rotation. 
The final rotation rate depends sensitively on how high an accretion rate the accretor can reach, and on how accretion ends, i.e., more gradually or more abruptly. If mass accretion rate declines more gradually, as in rotation-limited models, the star has more time to contract before accretion ceases and thus is closer to TE at the end of accretion.
Conversely, if accretion persists at a high rate until it ends more abruptly, for instance, in cases where rotation continues beyond the critical limit until the rapid decline of the mass-transfer rate, or when mass loss occurs through L2 overflow, the star has insufficient time to contract, leading to a greater departure from TE and a stronger subsequent reduction in fractional critical rotation.
Furthermore, as shown in Appendix\,\ref{app_sec:G}, inefficient internal angular momentum transport (e.g., in the absence of the Tayler–Spruit dynamo) can further enhance this reduction through internal angular-momentum redistribution.
More studies addressing the uncertainties in binary evolution are required to fully understand the post-accretion rotation rates of mass gainers.

\subsubsection{Case BB mass transfer}
It is important to note that in all of our binary models, the accretor undergoes a second phase of mass transfer (Case BB) due to the expansion of the donor star after core helium exhaustion. During this phase, the accretor again reaches critical rotation but shows no significant spin-down. This outcome is consistent with our conclusion, because Case BB mass transfer proceeds at much lower rates, allowing the accretor to remain close to TE.
The interval between Case B and Case BB mass transfer is short ($\leq$1\,Myr). Consequently, when Case BB mass transfer occurs, the accretor spends only a brief time rotating sub-critically after regaining TE from the preceding Case B mass transfer, before quickly returning to critical rotation during Case BB mass transfer.

However, Case BB mass transfer does not occur in all binaries. As discussed by \cite{1998brbi.book.....V}, stellar winds have a strong impact on the occurrence of Case BB mass transfer. If the stellar wind is sufficiently strong to remove the remaining hydrogen envelope of the donor star, no significant expansion is expected during later evolutionary stages, thereby suppressing Case BB mass transfer (see also \citealt{2018A&A...615A..78G,2020A&A...637A...6L}). In particular, the accretor in the binary model in \cite{2021ApJ...923..277R} does not undergo Case BB mass transfer.
%Case BB mass transfer will also not occur in binaries where the main-sequence accretor exhausts its central hydrogen before the donor depletes its central helium. 
In such systems, the spin-down that follows Case B mass transfer remains the decisive factor in setting the rotation of post-accretion stars. Their subsequent rotational evolution is then expected to resemble that of single stars, namely, the fractional critical rotation rate gradually increases as they approach the end of the main sequence \citep{2008A&A...478..467E,2020A&A...633A.165H}.

It is worth noting that a temporary detached phase during Case B mass transfer has been identified in detailed binary evolution models \citep{2022A&A...662A..56K}. In such binaries, although the accretor may spin down to sub-critical rotation after the first Case B mass-transfer episode, it can reach critical rotation again during a second stage of Case B mass transfer. Because this second stage proceeds at a much lower mass-transfer rate, the accretor may remain close to TE during this phase and stay near critical rotation afterward. This behavior is similar to that found in Case BB mass transfer. Such a temporary detached phase does not occur in our five binary models. A detailed investigation of the conditions under which this phase arises is beyond the scope of this study.

Finally, the extent of rotational reduction after Case B can influence the accretor’s surface composition following Case BB (see Fig.\,\ref{app_fig:binary_app} for details). A more slowly rotating accretor after Case B mass transfer can accrete more helium- and nitrogen-rich material during Case BB mass transfer. Thus, understanding rotational evolution during and after thermal-timescale accretion is critical for interpreting the surface abundances and rotational states of post-interaction stars in observations.

\subsection{Rotation of main-sequence merger products}
Our study also has implications for the rotation of main-sequence merger products. According to the simulations by \citet{2019Natur.574..211S}, the merger of two main-sequence stars results in an overheated product that is initially out of TE. They modelled the long-term evolution of the merger product with the 1D code MESA and found that the merger product spins down from critical rotation immediately after coalescence to about 10\% of the critical value by the time TE is restored. However, this conclusion is based on a single model. A comprehensive, systematic exploration of the rotational evolution of MS merger products is still lacking. 

Our results suggest that massive main-sequence merger products are likewise expected to experience a reduction in fractional critical rotation during thermal relaxation, although their internal structures may differ from those of our accreting models.
As long as the merger product possesses an expanded radiative envelope due to excess energy, a decrease in its fractional critical rotation should occur as it contracts toward TE.
The magnitude of this decrease likely depends on the degree of thermal disequilibrium at formation of the merger product and may be influenced by factors such as the progenitor mass ratio and on how the merger proceeds \citep[see][]{2024A&A...682A.169H}.

\subsection{Rotation of newly formed stars following accretion}
Our results may also provide insights into the spin evolution of newly formed massive stars. \citet{2017A&A...602A..17H} studied the rotational evolution of pre-main-sequence stars both during the accretion phase and in the subsequent relaxation phase after accretion ceased. They found that an initially fully convective low-mass pre-main-sequence seed can 
undergo substantial swelling during accretion as it gains mass and becomes fully radiative. Once the star becomes radiative, its structural response depends on the accretion rate: it may continue to expand for high rates (e.g., $\dot{M}=10^{-3}\,M_\odot\,\mathrm{yr}^{-1}$) or begin to contract for low rates (e.g., $\dot{M}=10^{-5}\,M_\odot\,\mathrm{yr}^{-1}$). If accretion stops before the star fully regains TE, it subsequently undergoes a thermal relaxation phase at the beginning of its main-sequence evolution.

\citet{2017A&A...602A..17H}  predicted that during this early thermal contraction phase, stars are expected to spin up.
This is because their models assume differential rotation: the initially convective star seed efficiently transports angular momentum inward, creating a rapidly rotating core. After the envelope becomes radiative, angular momentum transport is less efficient, so the surface lags behind. Once accretion ends, angular momentum from the core is gradually redistributed outward, spinning up the surface. However, they also tested a model with a solid-body rotation profile at the end of accretion, in which case the star exhibited moderate spin-down during the subsequent contraction phase in their early main-sequence evolution, which is consistent with our results.

Therefore, although protostellar accretion can also drive stars out of TE, the subsequent spin evolution may depend on both the accretion and angular momentum transport history. Unlike main-sequence accretors, pre-main-sequence stars evolve from fully convective to fully radiative structures. In massive stars, accretion may even continue after central hydrogen ignition, when a convective core has already formed. These structural differences may lead to distinct spin-evolution outcomes compared to our accreting models. Nevertheless, as shown by \citet{2017A&A...602A..17H}, if the rotational profile resembles that of our models, i.e., closer to solid-body rotation rather than differential rotation with a rapidly rotating core, a similar spin-down can occur during thermal relaxation. Clarifying how the rotation of newly formed stars connects to their thermal response after accretion remains an important open question for future study.

\subsection{Comparison with observations}
As discussed in Section\,\ref{sec:discussion1}, we can not reliably predict the post-mass-transfer rotation rates of stars, owing to major uncertainties in the physics of binary interaction. Nevertheless, our results suggest that such stars may exhibit a wide distribution of spin rates, ranging from slow rotation to near-critical values. Observations of systems known to have undergone binary interaction indeed provide evidence for sub-critical rotation in some cases.

\cite{2017MNRAS.464.2066S} measured the spin rates of eight O-type stars in WR+O binaries, and found that, on average, their equatorial rotational velocities are approximately 65\% of the critical values, significantly below critical rotation. Notably, at least two of these systems are long-period binaries where tidal forces are too weak to account for the observed sub-critical rotation. \cite{2016A&A...591A..22S} and \cite{2020MNRAS.492.4430S} also found that the O-type stars in WR+O binaries in the Magellanic Clouds rotate well below their critical velocities, even though stellar winds in these systems are not expected to be strong enough to cause significant spin-down.
%In particular, the system SMC AB8 provides compelling evidence for this:  the O-type star shows clear signs of rejuvenation, indicating that it has undergone a past mass transfer episode. Despite this, it rotates at only about 20–30\% of its critical velocity. Moreover, the binary’s long orbital period rules out tidal interaction as a viable mechanism for spin-down.
These findings are consistent with our prediction that post-interaction accretors can rotate below critical velocity. 

In Be+sdO candidate binaries, the Be stars are consistently measured as rapid rotators, but with a broad range of projected rotational velocities, from $250\,\kms$ up to $440\,\kms$ (see Table 2 in \citealt{2018ApJ...853..156W} and references therein). In addition, the O-type companion in the confirmed dormant black-hole binary VFTS 243 in the LMC rotates at only 50-60\% of its critical velocity, i.e., significantly below the critical limit. Recently, \cite{2025MNRAS.540.3523B} reported that SMC OB stars in SB1 systems also exhibit a wide range of rotational velocities.

Rapidly rotating runaway OB stars are also thought to be products of binary evolution, originating when their binaries are disrupted by a supernova explosion of the companion star. \cite{2022A&A...668L...5S} reported that the rapidly rotating runaway OB stars in 30 Doradus span a broad range of projected rotational velocities, from $200\kms$ up to $600\kms$. \cite{2024ApJS..272...45G,2025arXiv251120566G} also reported sub-critically rotating runaway Galactic stars identified in the Large Sky Area Multi-Object Fiber Spectroscopic Telescope (LAMOST) survey, which are likely products of binary evolution.

However, it is also important to note that \cite{2018A&A...615A..65V} proposed that magnetic fields 
generated during mass transfer may contribute to spinning down the accretor to sub-critical rotation rates. For example, in the case of Plaskett’s star, the mass gainer has been spun up through accretion but remains below critical rotation. Notably, the gainer exhibits a strong surface magnetic field exceeding 2.85 kG, suggesting that magnetic braking may play a significant role in its post-accretion spin evolution. Recently, \citet{2025MNRAS.tmp.1430B} proposed that magnetically coupled winds and star-disc coupling can prevent the accretor from reaching critical rotation during the accretion phase.
However, strong magnetic fields have not been widely detected in post-accretion stars, leaving open the question of whether magnetic braking is a common mechanism or a rare exception in their spin evolution.

%(cwang: Observations of Be star rotation: The average rotation of Be stars is about 88\% of critical rotation in our Galaxy (Fremat et al. 2005). Significantly subcritical rotation speeds were found for early-type Be stars (Cranmer 2005). Are there any binaries among these subcritically rotating stars? )

%(cwang, suggested by Mike: rapidly rotating Bn stars: TESS has observed many non-Be fast rotators:https://arxiv.org/pdf/2407.08305. In addition, Britavskiy et al. 2023 have investigated binarity among fast-rotating O-type stars as part of the IACOB project. But their rotation threshold is defined as $v\mathrm{sini}\geq 200\,\mathrm{km\,s^{-1}}$ and it remains unclear how close these stars are to their critical rotation velocities. This aspect needs further investigation.)
%===================================================================================================
\section{Concluding remarks}\label{sec:conclusion}
%===================================================================================================
In this study, we investigate the spin evolution of stars during the thermal relaxation phase that follows thermal-timescale mass transfer. We construct several sets of single-star models with different accretion prescriptions. All models converge on the same conclusion: stars spin down during thermal relaxation. The further the star departs from TE at the end of accretion, the stronger the subsequent spin-down.

The physical origin of this spin-down lies in the non-homologous structural response to accretion and thermal relaxation. In both phases, the tenuous envelope undergoes substantial changes, while the dense core remains largely unaffected. During accretion, the inflated envelope can absorb only a small amount of angular momentum, just enough to keep the expanded star rotating near the critical limit. After accretion ceases, contraction of the envelope has little effect on the stellar moment of inertia or surface angular velocity. However, as the stellar radius decreases, the critical angular velocity increases, causing the star to spin down from critical to sub-critical rotation. In cases of very high accretion rates, where rigid rotation cannot be maintained, additional spin-down occurs because the surface angular velocity itself decreases during thermal relaxation as angular momentum is transported inward. 
%Ultimately, the fundamental reason for spin-down is that the small angular momentum incorporated in the inflated envelope during accretion is insufficient to sustain critical rotation once the star contracts back into TE.

To test this conclusion in more realistic conditions, we compute five detailed binary models with different assumptions about accretion physics and angular momentum transport. These models also show spin-down during thermal relaxation, following the same trend as our single-star models, thereby supporting the robustness of our results.

Nevertheless, the rotation of accretors in real binaries cannot be precisely predicted because of major uncertainties in binary evolution, most notably in angular momentum transport, both within the star and between the star and its accretion disk. A key implication of our results is that thermal-timescale accretion can leave stars rotating below critical. In some cases, the surface velocity may even fall below the Be-star threshold. This spin-down can explain the sub-critical rotators observed in post-interaction binaries. It is important to note that assuming strong core–envelope coupling yields only a lower limit on the spin-down; otherwise, additional spin-down will occur as angular momentum is gradually redistributed internally. To fully understand the post-mass-transfer spins of accretors, more comprehensive binary models, including those that allow accretion beyond standard rotation-limited prescriptions, are urgently needed.

If the star later undergoes Case BB mass transfer, it will be spun up to critical rotation again and may remain critical for the rest of its main-sequence lifetime, as Case BB transfer is usually milder and does not drive large deviations from TE. However, any prior spin-down during Case B transfer strongly influences both the amount and the composition of material accreted during Case BB, thereby shaping the surface chemical abundances of the accretor.

Our findings also provide insight into the rotation of main-sequence merger products and massive protostars. Previous studies have shown that both are out of TE immediately after their formation; our results suggest that they should likewise undergo spin-down, similar to post-accretion stars. This conclusion is consistent with \citet{2019Natur.574..211S}. We advocate more detailed studies that explicitly account for stellar rotation and the associated physical processes in order to achieve a deeper understanding of the spin evolution of these systems.

\begin{acknowledgements}
C.W. gratefully acknowledges valuable comments from Hugues Sana, Ylva G\"otberg, Luqian Wang, and Dandan Wei on observational constraints on stellar rotation, as well as insightful discussions with Fabian Schneider, Eva Laplace, Mathieu Renzo, Hongwei Ge, and Hailiang Chen on theoretical explanations. C.W., X.D.L. and X.T.X. thank the National Key Research and Development Program of China (2021YFA0718500), and the National Natural Science Foundation of China (NSFC) under grant Nos. 12041301 and 12121003. C.W. also acknowledges funding from the Netherlands Organisation for Scientific Research (NWO) as part of the Vidi research program BinWaves (project number 639.042.728, PI: de Mink).

\end{acknowledgements}

\begin{appendix} %First appendix

%===================================================================================================
\section{Correlation between spin-down and thermal disequilibrium}\label{app_sec:A}
In this section, we analytically derive the relationship between the decline of fractional critical rotation velocity during the thermal relaxation phase and the deviation from thermal equilibrium (TE) at the end of mass accretion. The critical angular velocity is defined as  
\begin{equation}\label{eq:1}
\omega_\mathrm{crit}=\sqrt{(1-\Gamma)\frac{GM}{R^3}},
\end{equation}
where $M$, $R$, $\Gamma$ are the stellar mass, equatorial radius and Eddington factor, respectively. The Eddington factor before mass transfer is typically small, for example, less than 0.3 for a $12\mso$ star, and increases rapidly during mass transfer, with higher accretion rates resulting in larger Eddington factors. In single-star model Set I, for instance, a $12\mso$ star model with an initial accretion rate of $4.0\times10^{-3}\,M_\odot\,\mathrm{yr^{-1}}$ attains an Eddington factor of about 0.7 at critical rotation, whereas the corresponding model with an initial accretion rate of $1.0\times10^{-3}\,M_\odot\,\mathrm{yr^{-1}}$ reaches only approximately 0.45 at the same stage.
We can express the ratio of fractional critical velocity at the two epochs as
\begin{equation}\label{eq:2}
\frac{\frac{\omega_\mathrm{TE}}{\omega_\mathrm{crit,TE}}}{\frac{\omega_\mathrm T}{\omega_\mathrm{crit,T}}}=\frac{\omega_\mathrm{TE}}{\omega_\mathrm T}\frac{\omega_\mathrm{crit,T}}{\omega_\mathrm{crit,TE}}\simeq\frac{\omega_\mathrm{TE}}{\omega_\mathrm{T}}\Big(\frac{R_\mathrm{TE}}{R_\mathrm T}\Big)^{\frac{3}{2}},
\end{equation}
where the subscripts T and TE refer to the epochs when accretion terminates and when TE is re-established, respectively. $\omega$ and $\omega_\mathrm{crit}$ denote the stellar surface angular velocity and the surface critical angular velocity, respectively. Here we assume that stellar mass loss is negligible during thermal relaxation. Our stellar models rotate at nearly critical velocity at the termination of mass accretion, i.e., $\frac{\omega_\mathrm T}{\omega_\mathrm{crit,T}}\simeq 1$. It is worth noting that in our detailed binary models, the accretors can rotate sub-critically before mass transfer fully ceases, owing to the decline in the mass-transfer rate during the late stages of interaction (see Section\,\ref{app_sec:E}). To maintain numerical stability, we adopt a practical critical limit corresponding to $\sim98\%$ of the true critical velocity. 
Substituting the above relation into Eq.\,\ref{eq:2}, we obtain
\begin{equation}\label{eq:3}
\frac{\omega_\mathrm{TE}}{\omega_\mathrm{crit,TE}}\simeq\frac{\omega_\mathrm{TE}}{\omega_\mathrm{T}}\Big(\frac{R_\mathrm{TE}}{R_\mathrm T}\Big)^{\frac{3}{2}}.
\end{equation}

For stars accreting at relatively low rates, where rigid rotation is maintained, we have $\omega_\mathrm{TE} \simeq \omega_\mathrm{T}$, and thus $\frac{\omega_\mathrm{TE}}{\omega_\mathrm{crit,TE}}\sim \Big(\frac{R_\mathrm{TE}}{R_\mathrm T}\Big)^{\frac{3}{2}}$. We need to emphasis that, due to changes in stellar mass and structure during accretion, $\omega_\mathrm{TE}$ may differ from $\omega_\mathrm{T}$ even under rigid rotation. This results in a deviation of these models from the linear relation with a slope of unity (black dashed line in Fig.\,\ref{fig:v_k}).
In contrast, for stars with very high accretion rates, where rigid rotation breaks down, $\omega_\mathrm{TE} < \omega_\mathrm{T}$, i.e., surface angular velocity decreases during thermal relaxation. Therefore, for the same degree of thermal disequilibrium (i.e., the same $R_\mathrm{TE}/R_\mathrm{T}$), these stars experience a stronger spin-down during the relaxation phase.

\section{Results for single-star models in Set I}\label{app_sec:B}
\setcounter{figure}{0}
\renewcommand\thesection{\Alph{section}}
\renewcommand{\thefigure}{\thesection.\arabic{figure}}
\makeatletter
\renewcommand{\theHfigure}{\thesection.\arabic{figure}} % Update hyperref anchors
\makeatother

\begin{figure*}
\includegraphics[width=\linewidth]{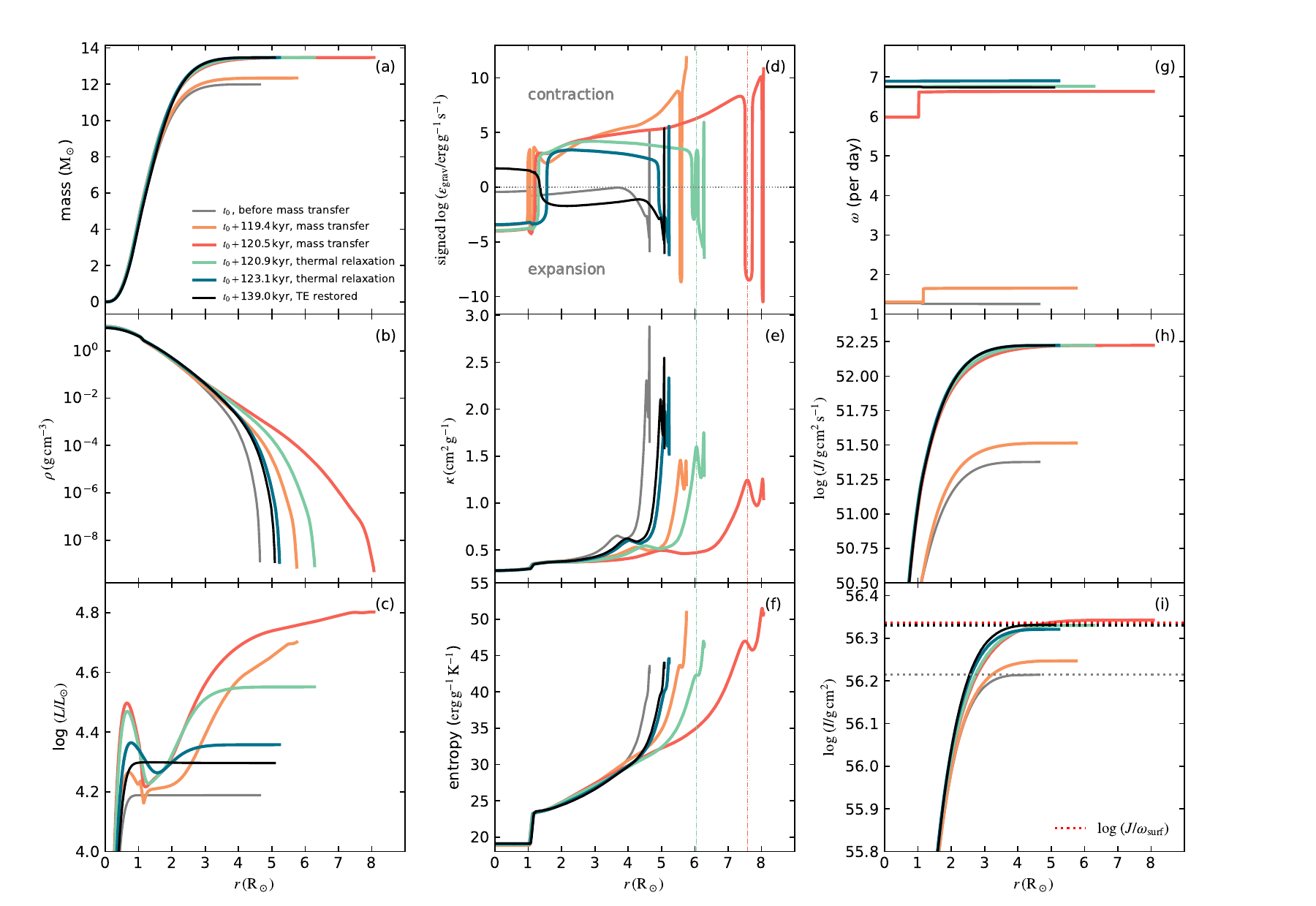}
\caption{Internal profiles of a $12\,M_\odot$ single-star model from Set I accreting at $1.0\times10^{-3}\,M_\odot\,\mathrm{yr^{-1}}$, shown as a function of radius coordinate. Colored lines correspond to six snapshots covering the phases before mass transfer, during mass transfer, during thermal relaxation, and after restoring thermal equilibrium. The reference time $t_0=10.06\,$Myr.
Panels: (a) enclosed mass; (b) density; (c) luminosity; (d) gravitational energy generation rate (positive=contraction, negative=expansion); (e) opacity; (f) entropy; The thin vertical dashed-dotted lines in Panels (d), (e) and (f) mark the radius of the iron opacity bump. (g) angular velocity; (h) enclosed angular momentum; (i) $\log I$ (solid lines) and $\log(J/\omega_\mathrm{surf})$ (horizontal dotted lines), where $J$, and $\omega_\mathrm{surf}$ denoting the total angular momentum, and surface angular velocity at the corresponding snapshots. The relation $I=J/\omega_\mathrm{surf}$ corresponds to rigid rotation, while $I>J/\omega_\mathrm{surf}$ indicates differential rotation, with the core rotating more slowly than the envelope. }
\label{fig:12Msun_profile}

\end{figure*}

\begin{figure*}
\includegraphics[width=\linewidth]{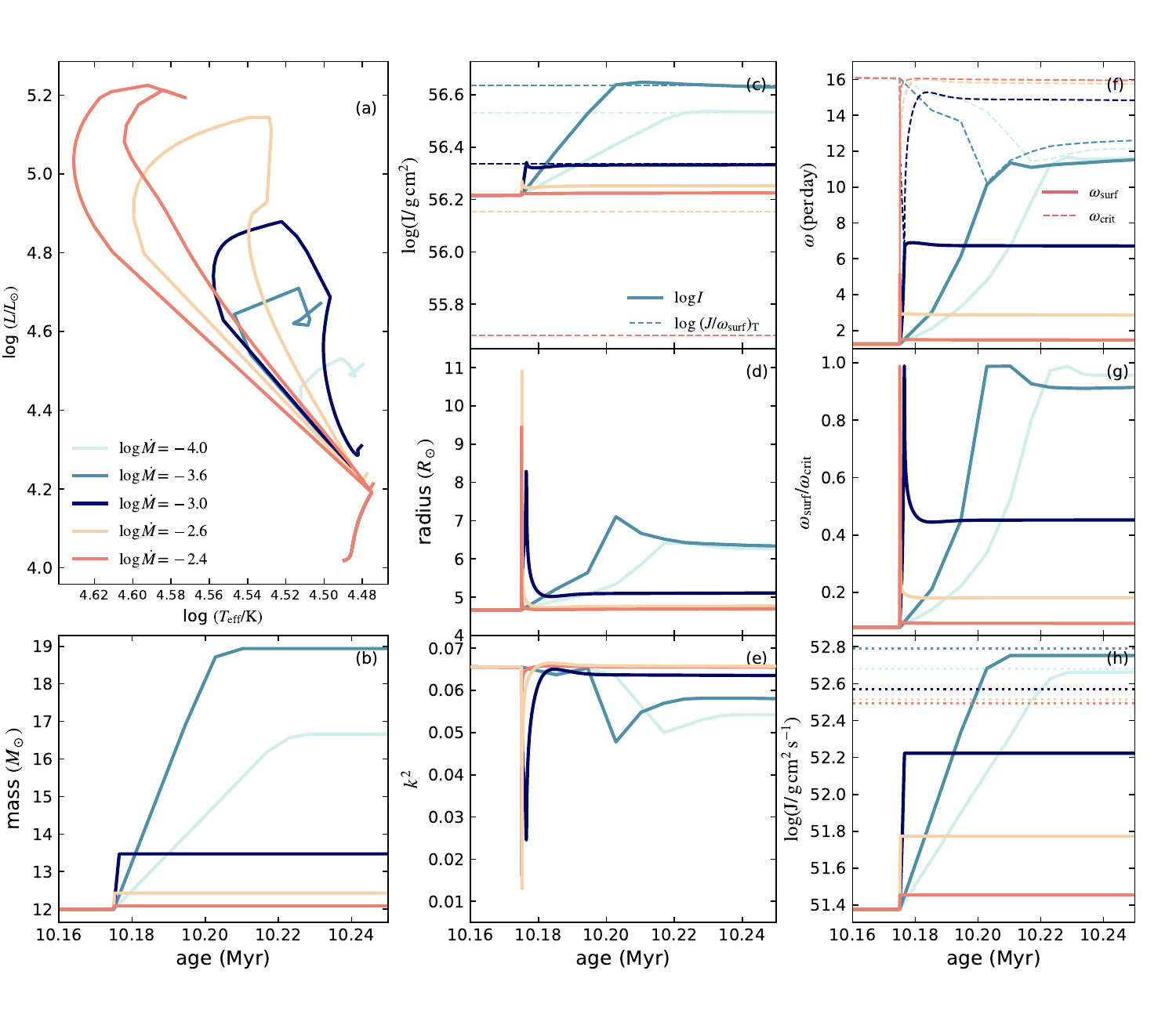}
\caption{Evolution of a 12$\mso$ single-star model accreting material at different constant rates. Each line color corresponds to a specific accretion rate, as indicated in the figure legend. Accretion begins when the central hydrogen abundance drops below 0.5 and terminates once the star reaches critical rotation. Panel (a): Evolutionary tracks in the Hertzsprung–Russell diagram. Panels (b)-(h): Time evolution of key stellar properties. (b) stellar mass; (c) moment of inertia $I$ (solid) and the logarithm of the ratio between the total angular momentum $J$ and the surface angular velocity $\omega$ at the termination of mass transfer (dashed); (d) stellar radius; (e) gyration constant $k^2$; (f) surface angular velocity (solid lines) and critical velocity (dashed lines); (g): ratio of surface angular velocity to critical velocity; (h): logarithm of the total angular momentum (solid line) and the total angular momentum required for the corresponding model to rotate at the critical velocities after thermal equilibrium is restored.}
\label{fig:stop_at_crit_rot_7figure}
\end{figure*}

\begin{figure*}
\includegraphics[width=\linewidth]{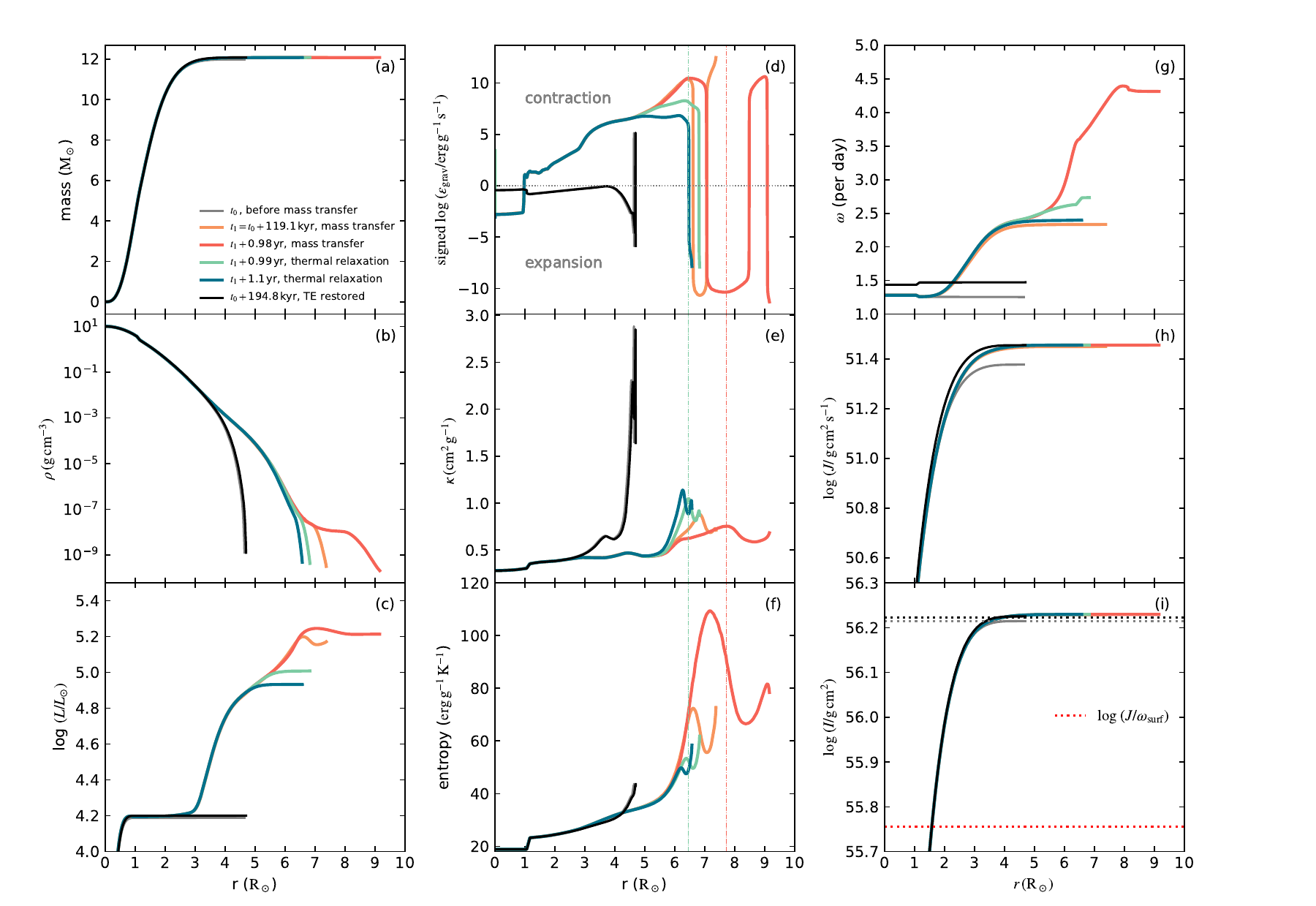}
\caption{Same as Fig.\,\ref{fig:12Msun_profile}, but for the $12\,M_\odot$ single-star model from Set I with an initial accretion rate of $4.0\times10^{-3}\,M_\odot\,\mathrm{yr^{-1}}$. Here the reference time $t_0=10.06\,$Myr.}
\label{fig:12Msun_profile_2.4}
\end{figure*}

\begin{figure*}
\includegraphics[width=\linewidth]{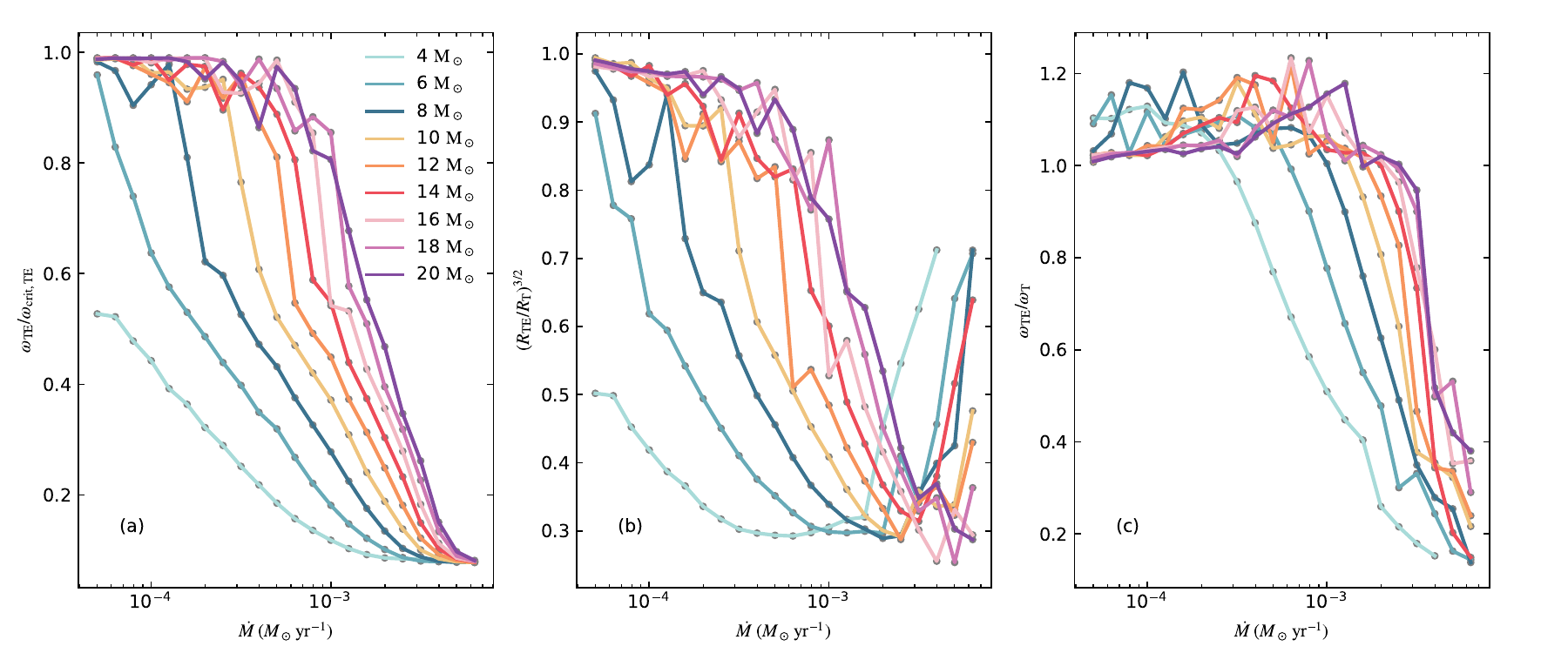}
\caption{Relationship between the mass accretion rate and (a) the ratio of the surface angular velocity to the critical value when the star regains thermal equilibrium (TE), (b) the deviation from thermal equilibrium characterized by the $(R_\mathrm{TE}/R_\mathrm{T})^{3/2}$, where $R_\mathrm{TE}$ and $R_\mathrm{T}$ denote the stellar radii at the times when TE is restored and when mass accretion terminates, respectively, and (c) the ratio between the surface angular velocity at the time of TE restoration ($\omega_\mathrm{TE}$) and that at the time of accretion termination ($\omega_\mathrm{T}$).
Different colors indicate models with different initial masses. Grey points along the lines represent individual models from Set I. }
\label{fig:stop_at_crit_rot}
\end{figure*}

%\begin{figure*}
%\includegraphics[width=\linewidth]{figures/modelI_J_mdot.pdf}
%\caption{Left: Solid lines show the ratio of total angular momentum at the termination of mass transfer to that before mass transfer (MT), as a function of accretion rate. Dotted lines show the ratio between the total angular momentum required for a thermally equilibrated star to reach critical rotation and that before mass transfer. Here, subscripts T and TE denote quantities at the termination of mass transfer and after thermal equilibrium is restored, respectively. Different colors indicate models with different initial masses. Right: Ratio of $k^2_\mathrm{T}r^2_\mathrm{T}/k^2_\mathrm{TE}r^2_\mathrm{TE}$ versus accretion rate for the same set of models.}
%\label{app_fig:I_J_Mdot}
%\end{figure*}

In this section, we present additional analysis of the single-star models from Set I to better understand the stellar response during accretion and subsequent thermal relaxation. We begin by examining in detail the internal structural evolution of the model shown in Fig.\,\ref{fig:fig1} of the main text, using its internal profiles at six snapshots. The results are presented in Fig.\,\ref{fig:12Msun_profile}.

Once mass transfer begins, the star is compressed by the newly accreted material and readjusts to a new hydrostatic equilibrium. This compression increases the central temperature and density of the star, enhancing the nuclear burning rate and raising the core luminosity (Panel c). Meanwhile, the envelope also experiences compression, leading to local contraction and the release of gravitational energy, which increases the luminosity in the outer layers. The dip in luminosity between the core and envelope marks the transition between the inner expanding layers, which absorb energy, and the outer contracting layers, which release energy (see Panel d for which layers experience expansion and contraction).

Because the accretion timescale is shorter than the thermal timescale, the star departs from TE, and the excess energy cannot be 
efficiently transported outward. The convective core is efficient at transporting energy and therefore undergoes only minor changes. This can be seen in Panels e and f that the radius of the convective core (approximately 1.1$\mathrm{R_\odot}$) changes very little during accretion. In contrast, the radiative envelope, especially regions with opacity bumps (notably the iron and helium opacity peaks), is inefficient in transporting energy. As a result, energy accumulates in these zones, causing them to expand significantly. This is evident by comparing Panels d and e (see the layers marked by dashed–dotted red vertical lines) that during accretion, the radius of the opacity peak coincides with the region of maximum gravitational energy absorption rate caused by expansion. 
%To maintain hydrostatic equilibrium, the surrounding layers contract in response (seen as alternating up and down features in the signed $\epsilon_\mathrm{grav}$ in Panel d).

At a low accretion rate of $1.0\times10^{-3}\,\mathrm{M_\odot\,yr^{-1}}$ for a $12\,M_\odot$ star, the star can accrete a notable amount of mass (about $2\,M_\odot$, Panel a) and angular momentum (Panel h) before reaching critical rotation. The accreted mass compresses the original material into deeper layers during accretion, as shown in Panel a by the leftward shift of the mass profile (i.e., smaller radii encompass more mass). The angular momentum is transported inwards, as illustrated by a similar shift in its profile. This inward transport of angular momentum keeps the star rotating nearly rigidly during accretion (Panel g). The moment of inertia also increases, with its profile shifting leftward. The largest changes in these profiles occur in layers with radii between roughly 2 and 4\,$R_\odot$, while the inflated outer envelope contributes negligibly to the mass, total angular momentum, and moment of inertia.

After accretion stops, the envelope remains overheated and overexpanded, with higher entropy (Panel f) than in a TE configuration. The star therefore contracts to restore TE. This contraction is non-homologous: the outer envelope contracts substantially, while the core expands slightly to adjust to the new equilibrium structure set by the increased total mass (e.g., Panels e and f). Because the tenuous envelope contains very little mass or moment of inertia, its large large change produces only a minor effect on the total stellar moment of inertia during thermal relaxation (Panel h). 
Once the excess energy is radiated away and the luminosity balances the nuclear burning rate, the star settles into TE (Panel c).

Next we explore the evolution of other models in Set I to study the impact of accretion rate on the spin evolution of post-accretion stars. Figure \ref{fig:stop_at_crit_rot_7figure} shows the evolution of $12\,\mso$ models under five different accretion rates. Panel (g) clearly demonstrates that the higher the accretion rate, the lower fractional critical velocity the star has after regaining TE. This is because, in general, a higher accretion rate leads to a larger expansion, and therefore, a larger deviation from TE during accretion (Panel d). 
This behavior is expected, as a higher accretion rate implies a larger disparity between the accretion and thermal timescales.
The exception is the model with a high accretion rate of $\log (\dot{M}/\mathrm{M_\odot\,yr^{-1}}) = -2.4$ (i.e., $\dot{M}\approx4.0\times10^{-3}\,\mathrm{M_\odot\,yr^{-1}}$), whose internal profiles at six snapshots, covering the phases before mass transfer, during mass transfer, during thermal relaxation and regaining TE, are shown in Fig.\,\ref{fig:12Msun_profile_2.4}. Panel g in Fig.\,\ref{fig:12Msun_profile_2.4} shows that at such a high rate, the star can no longer maintain rigid rotation during accretion. 
The stellar envelope rotates faster than the core, and the surface layers reach critical rotation more easily than in the rigid-rotation case. Consequently, the star does not need to expand as much as in lower-accretion-rate models to reach critical rotation.
Although this model does not exhibit the largest radial expansion, the decrease in surface angular velocity caused by angular momentum transport from the envelope to the core during thermal relaxation results in the strongest spin-down among the five models shown in Fig.\,\ref{fig:stop_at_crit_rot_7figure}.

With higher accretion rates, the accretor expands more rapidly and substantially, so its surface reaches critical rotation after accreting only a small amount of mass and angular momentum. Because the accreted mass is minimal, the luminosity from nuclear burning in the core changes little, while the luminosity in the outer layers increases significantly due to accretion luminosity (Panel c in Fig.\,\ref{fig:12Msun_profile_2.4}). In contrast, lower accretion rates result in longer accretion duration before the star reaches critical rotation, leading to larger total accreted mass and angular momentum (Panels b and h in Fig.\,\ref{fig:stop_at_crit_rot_7figure}). These results demonstrate that the degree of spin-down during thermal relaxation is not determined by the total amount of mass or angular momentum accreted, but rather by how fast mass and angular momentum are accreted. 
It is worth noting that at very low rates (e.g., $\log (\dot{M}/\mathrm{M_\odot\,yr^{-1}}) = -4.0$), the total accreted mass and angular momentum are actually smaller than those at a slightly higher rate (e.g., $\log (\dot{M}/\mathrm{M_\odot\,yr^{-1}}) = -3.6$), reflecting the interplay between accretion rate and duration.

Finally, we emphasize that the moment of inertia remains nearly constant during thermal relaxation for all models in Fig.\,\ref{fig:stop_at_crit_rot_7figure} (Panel c). As discussed in the main text, this is because the contraction of the tenuous, inflated envelope has negligible impact on the overall mass and angular momentum distribution.
For $\log (\dot{M}/\mathrm{M_\odot\,yr^{-1}}) \gtrsim -2.6$, the total moment of inertia exceeds the ratio between the total angular momentum and the surface angular velocity at the termination of mass transfer, indicating that the star rotates differentially at that point, with the core rotating more slowly than the envelope.

%The most prominent difference from the low accretion rate case lies in the angular velocity profile (Panel g). At high accretion rates, angular momentum cannot be efficiently transported from the surface to the core, leading to differential rotation. The core rotation remains almost unchanged during both accretion and thermal relaxation, while the envelope rotation evolves significantly. After the star regains TE, it returns to near-rigid rotation, with a slightly faster-rotating core and a substantially slower-rotating envelope compared to the end of accretion. In contrast, for the lower accretion rate case, the longer accretion timescale allows angular momentum to be transported inward efficiently during accretion. As a result, the angular momentum profile at the end of accretion is already similar to that after TE is restored (Panel g in Fig.\,\ref{fig:12Msun_profile}).

In Figure\,\ref{fig:stop_at_crit_rot}, we examine how the mass accretion rate affects stellar spin evolution and thermal disequilibrium at the end of accretion for models of different masses in Set I.
All masses follow the general trend described above: higher accretion rates lead to stronger spin-down (i.e., smaller $\omega_\mathrm{TE}/\omega_\mathrm{crit,TE}$ values when TE is restored). At a fixed accretion rate, lower-mass stars experience more pronounced spin-down due to their longer thermal timescales. 
For low accretion rates, where rigid rotation is maintained, $\omega_\mathrm{TE}/\omega_\mathrm{T}$ remains close to unity or slightly above (due to a small decrease in the moment of inertia during contraction). In these cases, the spin-down is primarily driven by the decrease of stellar radius, and therefore, the increase of critical rotation velocity. In contrast, at high accretion rates where rigid rotation cannot be sustained, $\omega_\mathrm{TE}/\omega_\mathrm{T}$ drops below unity. Here, although the stellar expansion may be smaller than in low $\dot{M}$ cases, the decrease in surface angular velocity, caused by angular momentum transport from the envelope to the core, plays an important role in the overall spin-down. The transition between these two regimes depends on stellar mass, with more massive stars able to maintain rigid rotation up to higher accretion rates. At the highest accretion rates, all models converge to $\omega_\mathrm{TE}/\omega_\mathrm{crit,TE}\approx0.08$, corresponding to the adopted initial rotation velocity of $50\,\mathrm{km\,s^{-1}}$.

%In Fig.\,\ref{app_fig:I_J_Mdot}, we show the angular momentum accreted during the accretion phase and the change of $k^2R^2$ during the subsequent thermal relaxation for stellar models in Set I. For reference, the left panel also includes the total angular momentum required for a star to rotate at critical velocity once it has regained TE. The figure illustrates that, for a given stellar mass, higher accretion rates lead to less angular momentum being accreted and to a larger deficit relative to the amount required for a thermally relaxed star to rotate at critical velocity.

%The right panel shows that at lower accretion rates, the stellar moment of inertia remains nearly constant during thermal relaxation, whereas at higher accretion rates, it increases. As we discuss later in this section, this difference arises from the balance between the accretion timescale and the angular momentum transport timescale.

\section{Results for single-star models in Sets II and III}\label{app_sec:C}
\setcounter{figure}{0}
\renewcommand\thesection{\Alph{section}}
\renewcommand{\thefigure}{\thesection.\arabic{figure}}
\makeatletter
\renewcommand{\theHfigure}{\thesection.\arabic{figure}} % Update hyperref anchors
\makeatother
\begin{figure*}
\includegraphics[width=\linewidth]{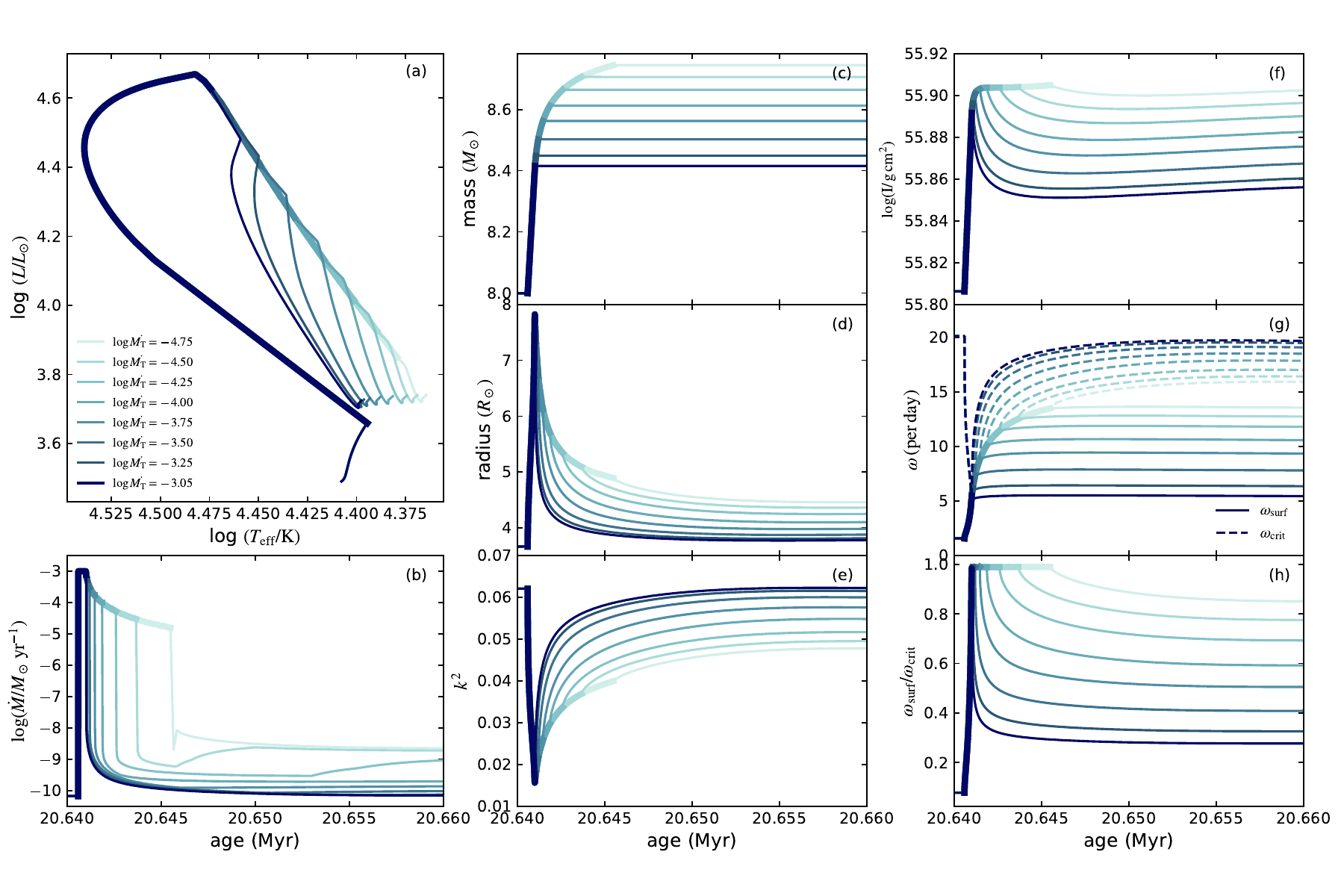}
\caption{Evolution of 8$\mso$ single-star models from Set II with an initial accretion rate of $1.0\times 10^{-3}\mso/ \mathrm{yr}$.  Accretion begins when the central hydrogen mass fraction reaches 0.5. After the star first attains critical rotation, the accretion rate is gradually reduced to keep the surface rotation just below the critical value, and is then terminated at different rates, as indicated by the colored lines corresponding to the values listed in the legend.
Panel (a) shows the evolution of the stellar model in the Hertzsprung-Russell diagram. Panels (b) to (h) present the time evolution of key stellar properties. (b) mass accretion rate; (c) stellar mass; (d) stellar radius; (e) gyration constant $k^2$; (f) moment of inertia; (g) surface angular velocity (solid lines) and critical rotation velocity (dashed lines); (h) ratio of rotational to critical angular velocity. In all panels, thick lines highlight the accretion stage.}
\label{fig:8Msun_main}
\end{figure*}

\begin{figure*}
\includegraphics[width=\linewidth]{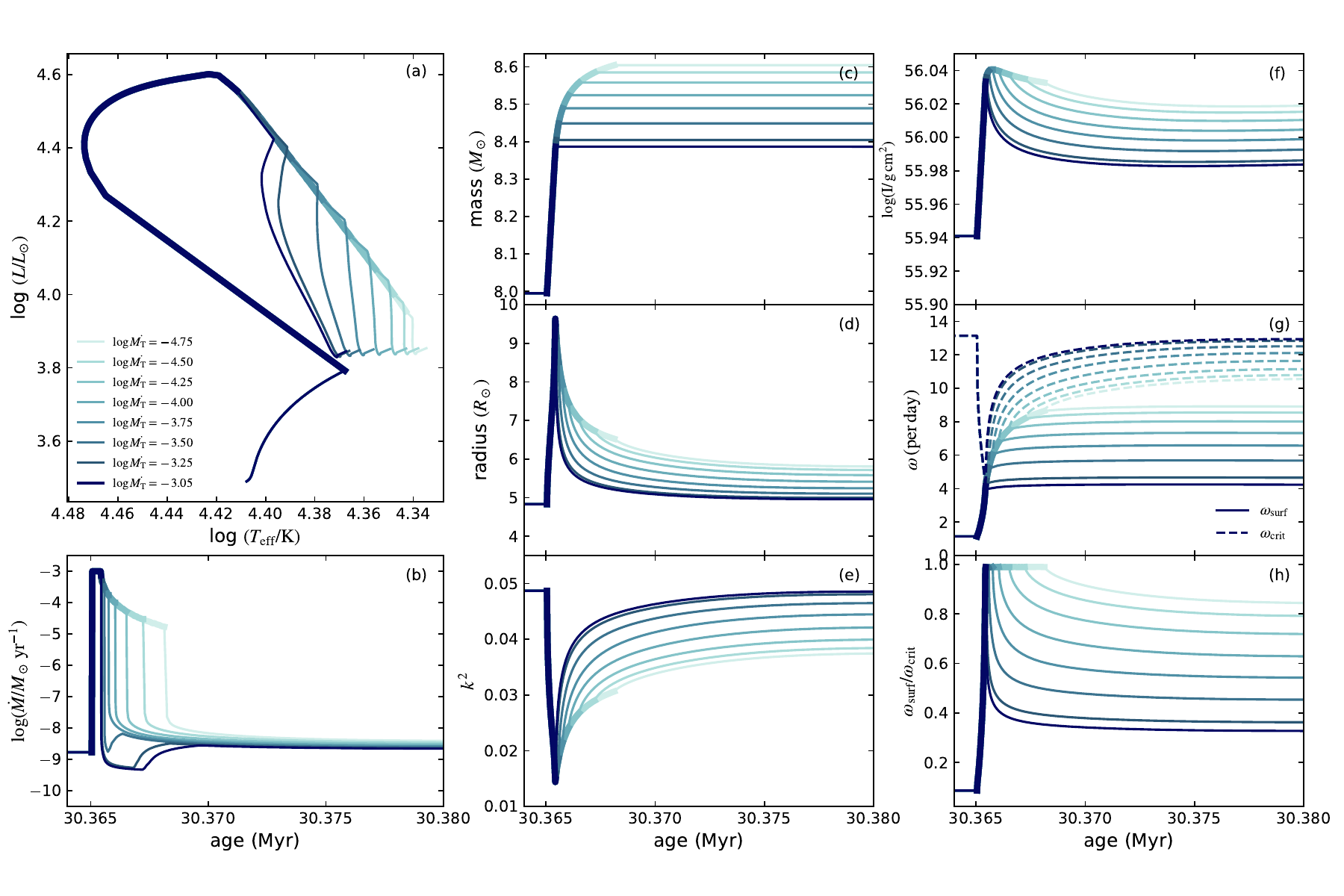}
\caption{Same as Fig.\,\ref{fig:8Msun_main}, but for single-star models in Set III, in which mass accretion happens when central hydrogen mass fraction reaches 0.3. The evolution stops when central hydrogen mass fraction drops to 0.28.}
\label{app_fig:8Msun_H03}
\end{figure*}

Single star models in Set II represent a more realistic scenario in which accretion continues at a reduced rate once the star reaches critical rotation. This setup accounts for the star’s structural response during mass accretion. In these models, thermal relaxation occurs concurrently with ongoing accretion, and the star is kept at the critical rotation limit until accretion is artificially stopped. By terminating accretion at different times, we effectively sample different degrees of deviation from TE: the later accretion stops, the closer the star is to regaining TE. Below, we examine whether these models exhibit similar spin-down behavior as observed in Set I.

The evolution of models in Set II is shown in Fig.\,\ref{fig:8Msun_main}. It can be seen that the earlier the accretion is stopped, the farther the star is from TE (Panel d), and the more significant the subsequent spin-down becomes (Panel h). This agrees with the findings from Set I. Similar as Set I models, although later termination (i.e., longer accretion duration) results in more mass being accreted (Panel c), the degree of spin-down is governed by how far the star is from TE, not by the total accreted mass.

The star’s final position in the Hertzsprung–Russell diagram after thermal adjustment is governed by two main factors: the total accreted mass (which determines the luminosity and the extent of rejuvenation) and the surface rotation rate. As shown in Panels c and h, the model with the longest accretion duration ends up with both the highest mass and the highest rotation rate. While the increase in mass tends to move the star toward higher effective temperatures and luminosities, rapid rotation counteracts this trend by lowering both quantities through centrifugal effects. Consequently, the model with the longest accretion duration exhibits a lower effective temperature and a similar luminosity compared to the model with the shortest accretion duration after regaining TE (Panel a).

To test whether the accretor’s evolutionary stage at the onset of mass accretion influences the thermal relaxation behavior, we construct Set III models, which follow the same accretion prescription and stopping criteria as models in Set II, but with accretion beginning when the central hydrogen abundance drops to 0.3 instead of 0.5. The evolution of this model set is shown in Figure\,\ref{app_fig:8Msun_H03}. Once again, we find that the strength of the spin-down correlates with the deviation from TE at the end of accretion, indicating that our conclusion is robust against the evolutionary stage of the accretor.

%===================================================================================================

\section{Results for single-star models in Set IV}\label{app_sec:D}

\setcounter{figure}{0}
\renewcommand\thesection{\Alph{section}}
\renewcommand{\thefigure}{\thesection.\arabic{figure}}
\makeatletter
\renewcommand{\theHfigure}{\thesection.\arabic{figure}} % Update hyperref anchors
\makeatother

\begin{figure*}
\includegraphics[width=\linewidth]{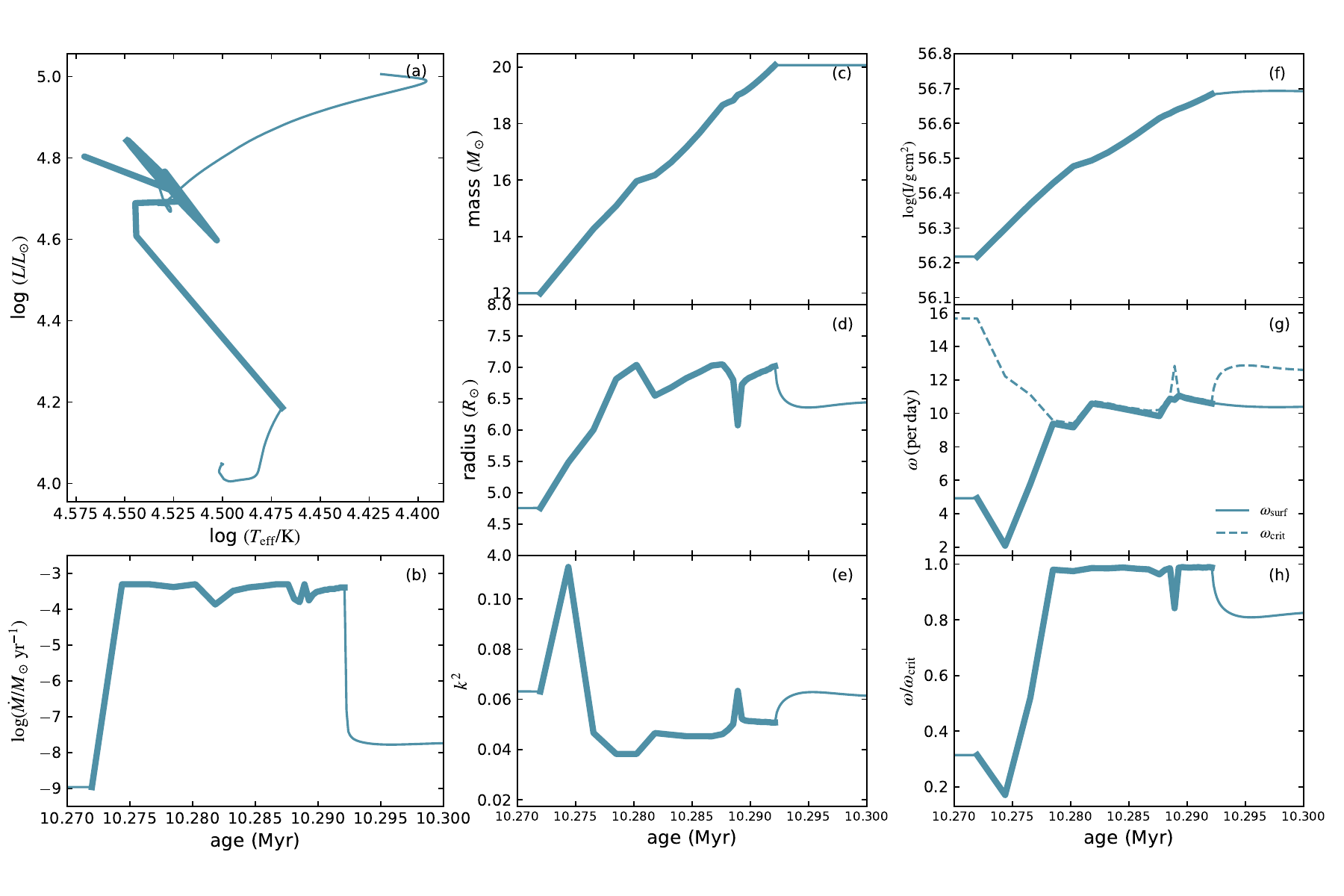}
\caption{Evolution of a 12$\mso$ single-star model from Set IV. Accretion begins when the central hydrogen mass fraction reaches 0.5 and proceeds at a constant rate of $5\times10^{-4}\,M_\odot\,\mathrm{yr^{-1}}$ until the stellar mass grows to $20\,\mso$. The specific angular momentum of the accreted material is artificially reduced by a factor of ten to enable continued accretion after the star first reaches critical rotation. The panels are the same as those in Fig.\,\ref{fig:8Msun_main}. The thick lines highlight the accretion phase.}
\label{app_fig:12_to_20_5d4}
\end{figure*}

\begin{figure*}
\includegraphics[width=\linewidth]{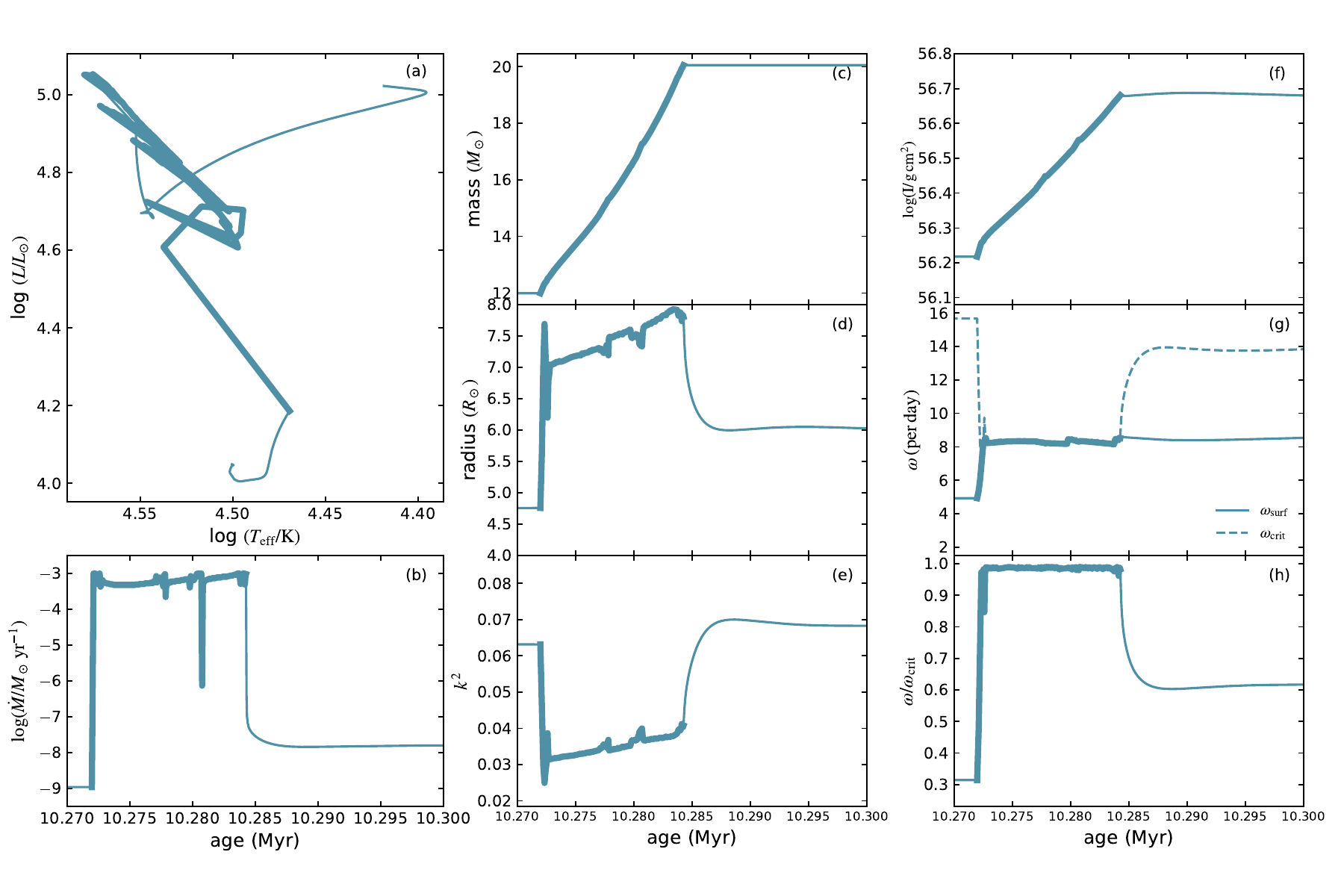}
\caption{As in Fig.\,\ref{app_fig:12_to_20_5d4}, but for the single-star model in Set IV, which accretes material at a constant rate of $1\times10^{-3}\,M_\odot\,\mathrm{yr^{-1}}$.}
\label{app_fig:12_to_20_1d3}
\end{figure*}

Recent studies of various types of binary systems suggest that accretion efficiencies may be significantly higher than those predicted by rotation-limited binary models \citep{2007ASPC..367..387P,2014ApJ...796...37S,2018A&A...615A..30S,2025arXiv250514780L,2025arXiv250323876X}. One proposed mechanism to achieve this is the outward transport of angular momentum  \citep{1991ApJ...370..604P,1991ApJ...370..597P,1991MNRAS.253...55C}, which enables continued mass accretion even after the accretor reaches critical rotation. Due to the lack of self-consistent detailed binary models that incorporate such processes, we construct simplified single-star models in Set IV to explore this scenario.

In Set IV, we compute the evolution of $12\mso$ stars accreting at constant rates of $5\times10^{-4}\,M_\odot\,\mathrm{yr^{-1}}$ and $1\times10^{-3}\,M_\odot\,\mathrm{yr^{-1}}$. Once the stars reach critical rotation, we artificially reduce the specific angular momentum of the accreted material by a factor of 10. While this adjustment is not based on a specific physical mechanism, it serves as a proxy to mimic the effect of enhanced angular momentum loss, enabling continued accretion at high rates without exceeding the critical rotation limit. Accretion is stopped once the stars reach a total mass of $20\mso$. The accretion efficiencies of these models are approximately 79\% and 65\%, respectively.

The results for models in Set IV are shown in Figs.\,\ref{app_fig:12_to_20_5d4} and \ref{app_fig:12_to_20_1d3}. In these models, the stars remain in an expanded configuration throughout accretion, with contraction occurring only once mass accretion is manually terminated. The subsequent contraction and spin-down then follow the same trend observed in the other single-star models (see Fig.\,\ref{fig:v_k}). This experiment demonstrates that the link between stellar spin-down and thermal disequilibrium is robust and does not depend on accretion prescription.

%As the star gains mass, it does not continue expanding monotonically from its previous radius. Instead, at each stage, the radius adjusts to the ongoing accretion rate relative to the star’s current mass. Once accretion ceases, the relation between spin-down and the deviation from TE remains consistent with our main conclusion.

\section{Results for binary models I, II and III}\label{app_sec:E}
\setcounter{figure}{0}
\renewcommand\thesection{\Alph{section}}
\renewcommand{\thefigure}{\thesection.\arabic{figure}}
\makeatletter
\renewcommand{\theHfigure}{\thesection.\arabic{figure}} % Update hyperref anchors
\makeatother

\begin{figure*}
\includegraphics[width=\linewidth]{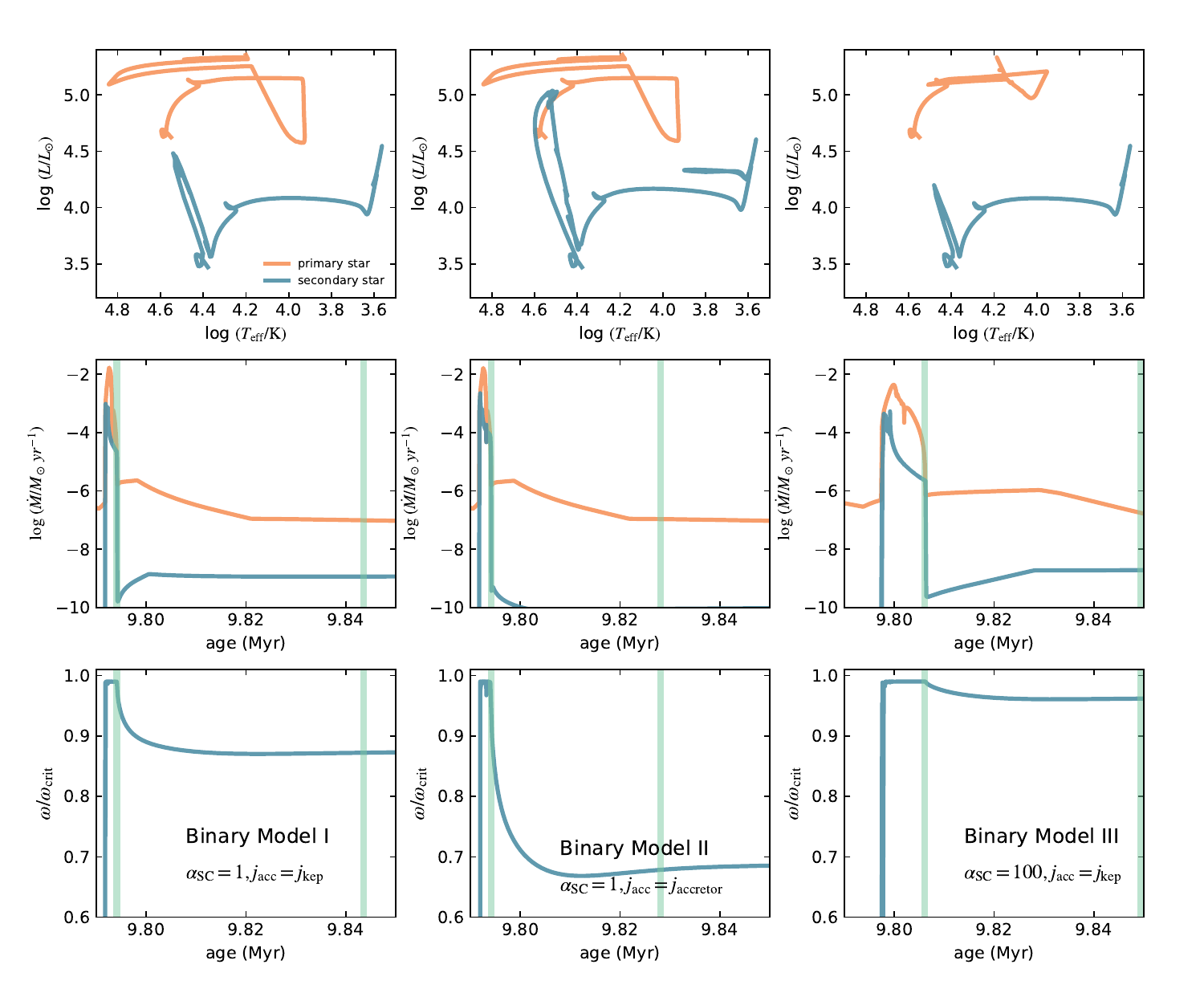}
\caption{Evolution of binary models I (left panels), II (middle panels) and III (right panels), with their initial parameters and physical assumptions listed in Table\,1. The top panels show the evolution of the primary (orange) and secondary (blue) stars in the Hertzsprung–Russell diagram. The middle panels display the evolution of the mass loss rate as a function of stellar age, using the same color scheme. The bottom panels show the time evolution of $\omega/\omega_\mathrm{crit}$ for the secondary star. The thick vertical cyan lines indicate, from left to right, the times when mass transfer ceases and when thermal equilibrium is restored.}
\label{app_fig:binary_main}
\end{figure*}

\begin{figure*}
\includegraphics[width=\linewidth]{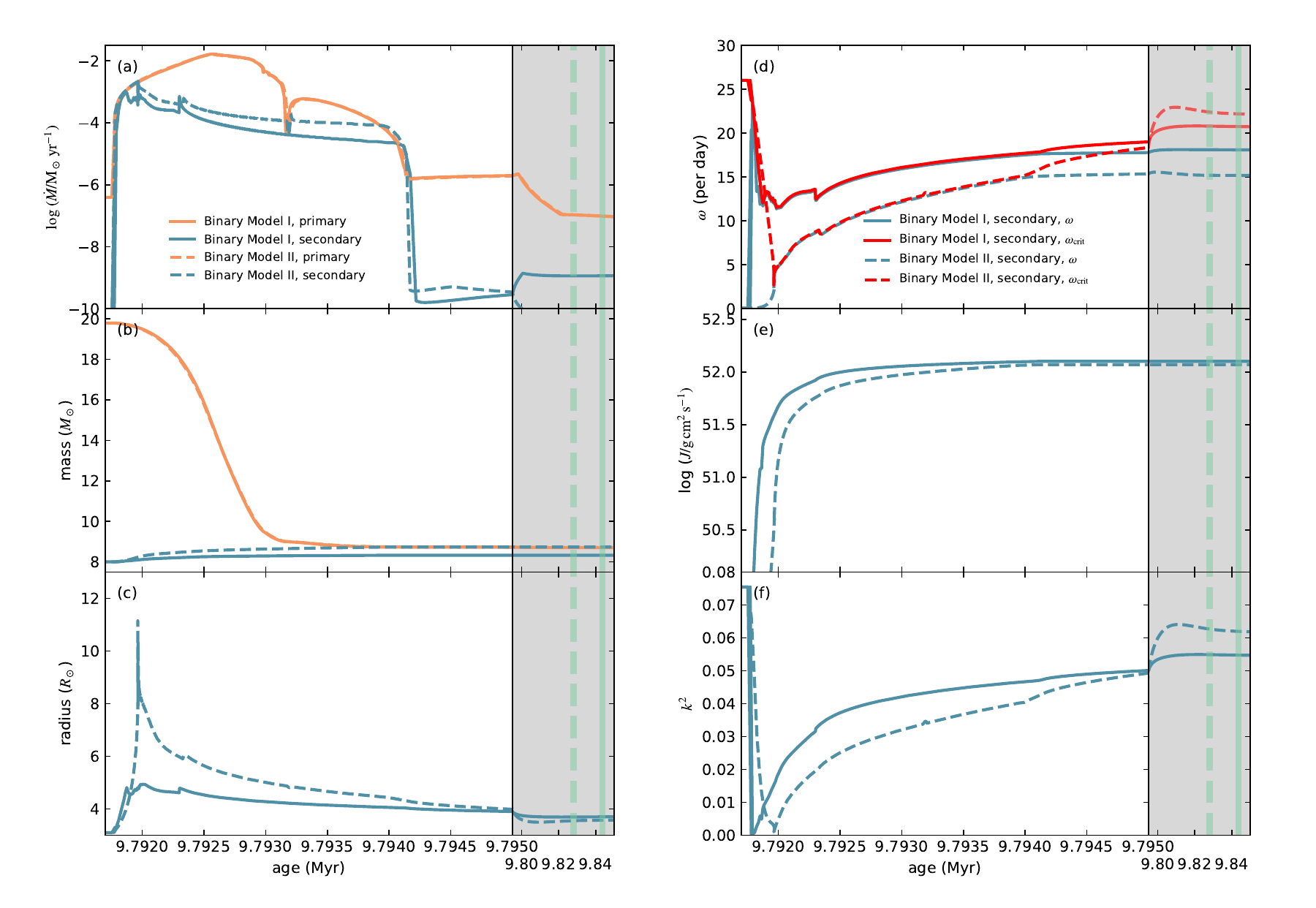}
\caption{Comparison of the evolution of binary models I and II during (white left region of each panel) and after (grey right region) Case B mass transfer. Each panel is divided such that the narrower right portion represents a longer timescale than the broader left portion. Panels (a) to (f) show the evolution of: (a) mass change rate; (b) stellar mass; (c) stellar radius; (d) stellar surface angular velocity and critical angular velocity; (e) total angular momentum; (f) gyration constant. Orange lines represent the evolution of the primary stars (solid for Model I, dashed for Model II), and blue lines represent the evolution of the secondary stars (with the same linestyles). In Panel (d), red lines denote the critical angular velocities of the secondary stars. 
In the grey region, the dashed and solid vertical cyan lines indicate the time when the secondary stars in Models II and I, respectively, re-establish thermal equilibrium.}
\label{app_fig:binary_app2}
\end{figure*}

\begin{figure}
\includegraphics[width=\linewidth]{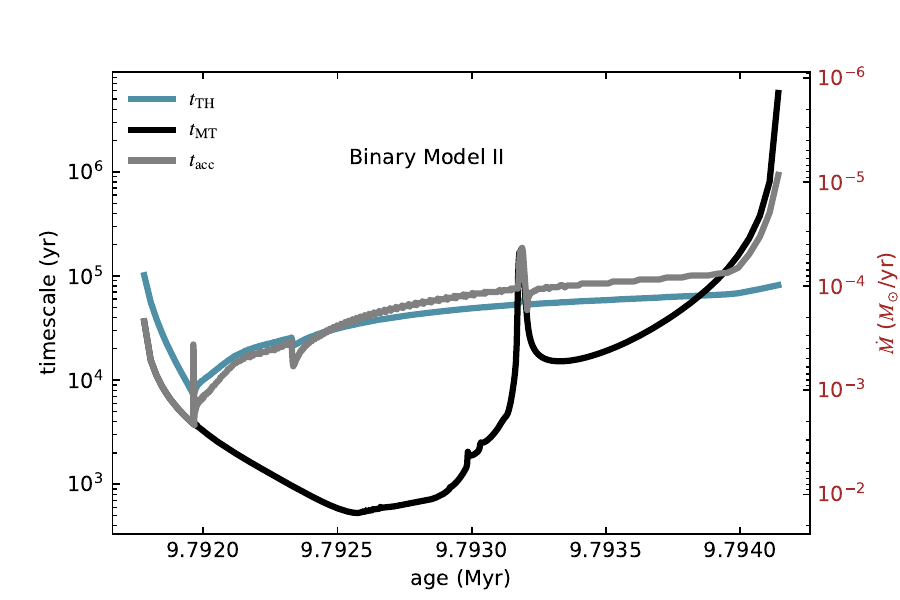}
\caption{Comparison of characteristic timescales during Case B mass transfer for the accretor in binary model II. The blue line shows the evolution of the accretor’s thermal timescale, while the grey line indicates the accretion timescale based on the actual accretion rate. The black line shows the accretion timescale assuming the accretion rate equals the mass-transfer rate. The right-hand y-axis gives the mass-accretion rate required for an 8$\mso$ star to reach an accretion timescale corresponding to the values on the left-hand axis.} 
\label{app_fig:binary_timescale}
\end{figure}

\begin{figure*}
\includegraphics[width=\linewidth]{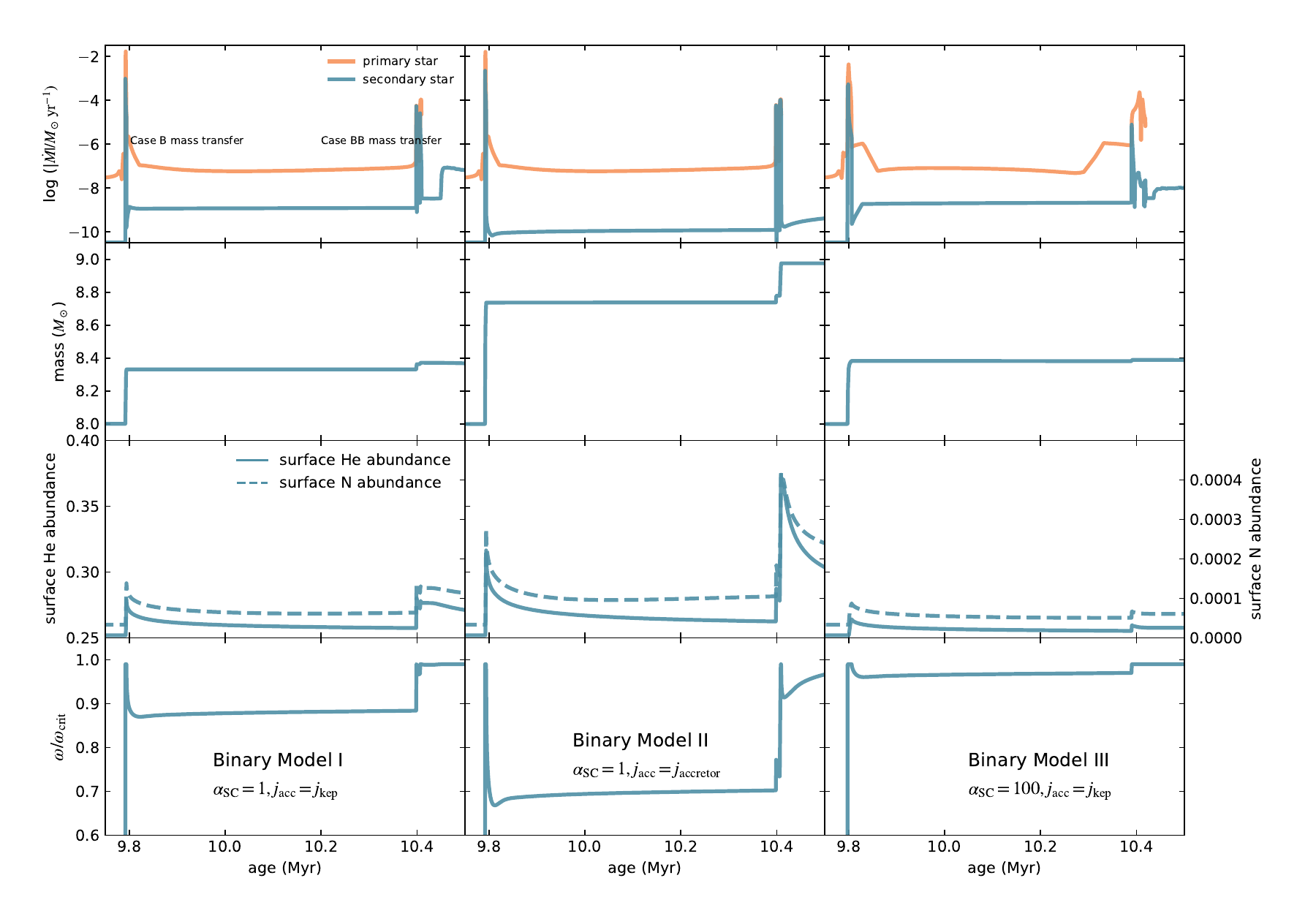}
\caption{Evolution of mass loss rate (top row), stellar mass (second row), surface chemical composition (third row) and rotational velocity $\omega/\omega_\mathrm{crit}$ (bottom row) as a function of age for binary models I (left), II (middle) and III (right). Orange and blue lines correspond to the primary and secondary stars, respectively. For the surface chemical composition, the helium mass fraction (blue solid lines) and nitrogen mass fraction (blue dashed lines) are shown for the secondary star.}
\label{app_fig:binary_app}
\end{figure*}

In this appendix, we show the evolution of binary models I, II and III, which adopt physics assumptions corresponding to those in 
\cite{2020ApJ...888L..12W}, \cite{2021ApJ...923..277R} and \cite{2022A&A...662A..56K}, respectively (see Section\,\ref{sec:method} for details). 
Figure\,\ref{app_fig:binary_main} highlights the evolution of the binaries in the Hertzsprung-Russell diagram, along with mass transfer and spin evolution during accretion and the subsequent thermal relaxation phase. 
In Fig.\,\ref{app_fig:binary_app2}, we directly compare the properties of binary models I and II.

The donor properties are identical in Models I and II, so their overall mass-transfer histories are similar. The difference lies in the assumptions for the specific angular momentum of the accreted material, which lead to distinct accretion histories. Model II assumes a lower specific angular momentum for the accreted material. Consequently, the accretor reaches critical rotation later, allowing more mass to be accreted at higher instantaneous rates before hitting the critical limit (see Fig.\,\ref{app_fig:binary_app2}). This produces a larger maximum radius and luminosity compared to Model I. After critical rotation is reached, the accretion rate in Model II declines more gradually, again owing to the assumed lower specific angular momentum of the incoming material. We find that the accretor in Model II drops below critical rotation even before mass transfer fully ends, as the declining accretion rate becomes insufficient to sustain critical rotation. Together, these effects produce a larger deviation from TE at the end of mass accretion, and therefore a stronger spin-down during thermal relaxation in Model II. The relationship between $\omega_\mathrm{TE}/\omega_{\mathrm{TE,crit}}$  and $(R_\mathrm{TE}/R_\mathrm{T})^{3/2}$ remains consistent with the trends predicted by our single-star models (see Fig.\,\ref{fig:v_k}).

In Model III, mass transfer occurs over a longer timescale and at a lower rate than in the other two models, owing to the donor’s distinct response under the assumption of high semi-convective mixing efficiency (see \citealt{2022A&A...662A..56K} for detailed discussion). The smaller deviation from TE at the point when the accretor reaches critical rotation (as indicated by the lower peak luminosity during accretion), combined with the longer contraction phase during accretion, leads to an overall smaller departure from equilibrium by the end of mass transfer. As a result, this model exhibits almost no spin-down. 

In these detailed binary evolution models, the accretor begins to contract after reaching critical rotation, even while accretion continues at a reduced rate. This indicates that the reduced accretion rate is no longer sufficient to drive the star further out of TE.
To illustrate this, we compute the accretion and thermal timescales for the accretor in binary Model II, following the prescriptions of \citet{2024ApJ...966L...7L}.
The thermal timescale is given by:
$$t_\mathrm{TH} = \frac{GM^2}{R_\mathrm{eff}(M)L},$$
where $M$ and $L$ denote the instantaneous stellar mass and luminosity. The effective radius is defined as
$$\frac{R_\mathrm{eff}(M)}{R_\odot}=2\Big(\frac{M}{M_\odot}\Big)^{0.22},$$
and represents an effective radius that characterizes the depth in the stellar interior from which gravitational energy is predominantly released. The mass-accretion timescale is expressed as
$$t_\mathrm{acc} = \frac{M}{\dot{M}_\mathrm{acc}},$$
where $\dot{M}_\mathrm{acc}$ is the instataneous accretion rate. 
To account for the possibility that accretion may continue at high rates even after the star reaches critical rotation, we also consider the extreme case where the accretion rate always equals the mass-transfer rate, yielding
$$t_\mathrm{MT} = \frac{M}{\dot{M}_\mathrm{MT}},$$
with $\dot{M}_\mathrm{MT}$ the mass transfer rate from the donor star. 

A comparison of the three characteristic timescales is shown in Fig.\,\ref{app_fig:binary_timescale}. It can be seen that once the star reaches critical rotation, the accretion timescale becomes comparable to, or longer than, the thermal timescale owing to the declining accretion rate. This is consistent with \citet{2024ApJ...966L...7L}, who found that stellar contraction begins when $t_\mathrm{TH}/t_\mathrm{acc} \approx 1$. 

Comparing $t_\mathrm{MT}$ and $t_\mathrm{acc}$ shows that thermal contraction inevitably occurs before mass transfer ends, even in the case of fully conservative mass transfer. This occurs because the mass-transfer rate declines rapidly during the late stages of the interaction, eventually making the two timescales comparable.
However, this final stage, during which the accretor must contract as the mass-transfer rate decreases, is very short ($<1000$\,yr), much shorter than the stellar thermal timescale (tens of thousands of years). Consequently, the star cannot fully regain TE by the end of mass transfer in the case of fully conservative mass transfer.
This analysis demonstrates that, in conservative mass transfer, the early, high-rate phase of mass transfer, rather than the short final low-rate phase, primarily determines how far the accretor departs from TE at the end of accretion, and thus governs the extent of its subsequent rotational reduction.

It is important to note that in all three binary models, the accretor undergoes a subsequent Case BB mass transfer triggered by the expansion of the donor star following core helium exhaustion (see Fig.\,\ref{app_fig:binary_app}). During this phase, the accretor once again reaches critical rotation but does not experience significant spin-down. This behavior is consistent with our findings from Case B mass transfer, where the degree of spin-down is closely linked to the thermal disequilibrium induced by rapid accretion. In Case BB, however, the mass transfer rates are substantially lower, allowing the accretor to remain much closer to TE, and therefore, no notable spin-down occurs. The efficiency of accretion during Case BB mass transfer is strongly influenced by the rotational state of the accretor at the end of the previous Case B mass transfer phase. Specifically, stronger spin-down during Case B mass transfer allows the accretor to accrete more mass in Case BB mass transfer, thereby altering its surface chemical composition. Among the three models, Model II exhibits the strongest surface enrichment in both helium and nitrogen. Compared to Model I, the nitrogen abundance increases by a factor of 2.09 ($2.3 \times 10^{-4}$ vs. $1.1 \times 10^{-4}$ at 10.5\,Myr after Case BB mass transfer), while the helium abundance shows a more modest increase of 1.11 (0.30 in Model II vs. 0.27 in Model I at 10.5\,Myr). In contrast, the accretor in Model III, which remains near critical rotation following Case B, accretes little to no material during Case BB, and therefore shows the least surface enrichment among the three.

%Two key relevant uncertainties in binary evolution affect the outcome. The first one is how quickly the accretion rate decreases (compare binary Models I and II), and second one is when accretion is ultimately terminated (compare Models I and III). Both factors influence how far the star is from TE when accretion ends, which in turn determines the extent of spin-down during thermal readjustment.

%\bibliography{sample63}{}
%\bibliographystyle{aasjournal}



\section{Results for binary model IV}\label{app_sec:F}
\setcounter{figure}{0}
\renewcommand\thesection{\Alph{section}}
\renewcommand{\thefigure}{\thesection.\arabic{figure}}
\makeatletter
\renewcommand{\theHfigure}{\thesection.\arabic{figure}} % Update hyperref anchors
\makeatother

\begin{figure*}
\includegraphics[width=\linewidth]{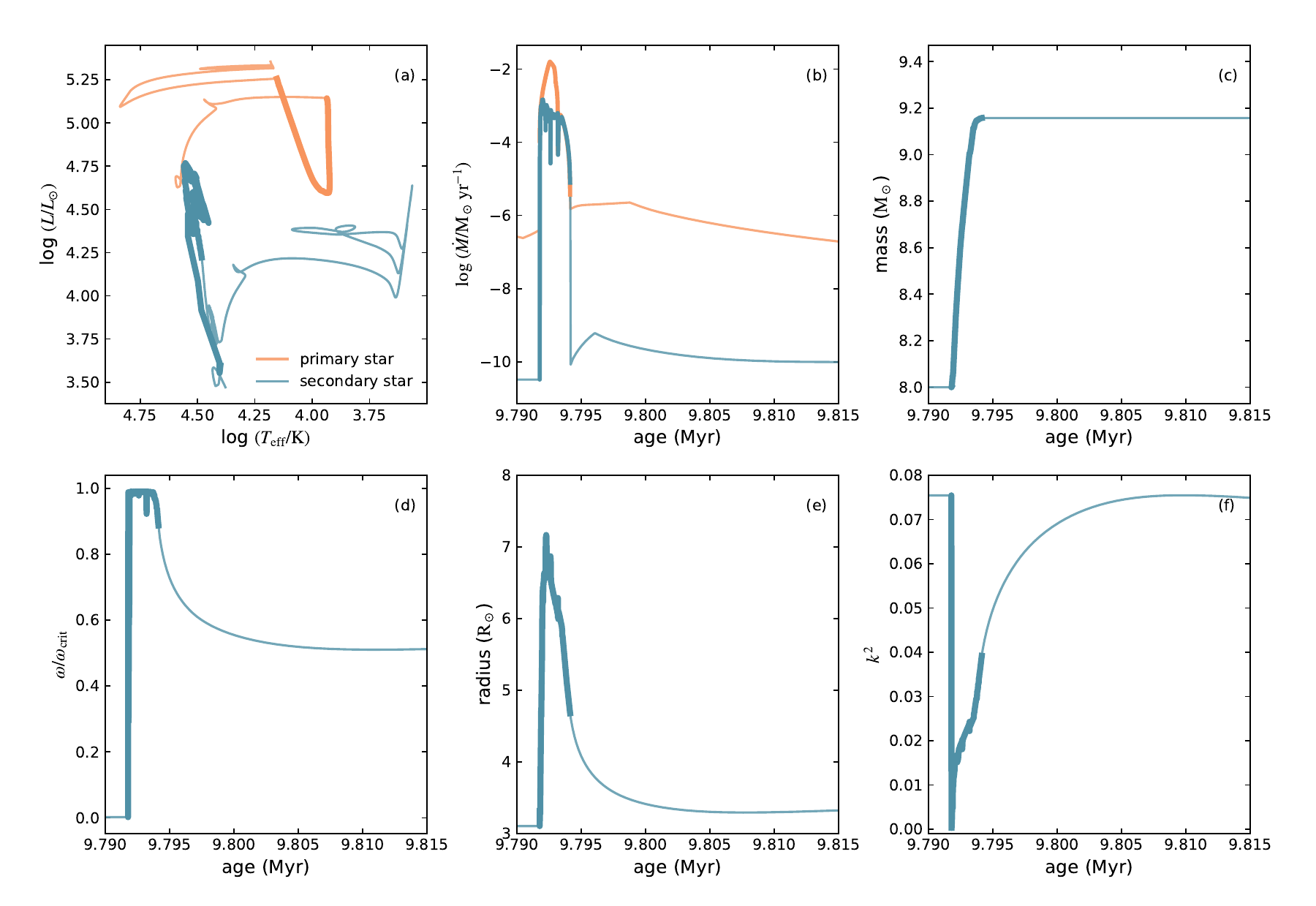}
\caption{Evolution of binary Model IV, in which the specific angular momentum of the accreted material is artificially reduced by a factor of ten once the accretor reaches critical rotation. Panel (a): evolution in the Hertzsprung–Russell diagram; (b) mass-transfer and accretion history; (c) stellar mass; (d): ratio of surface angular velocity to the critical velocity; (e): stellar radius; (f) gyration constant.}
\label{app_fig:binary4}
\end{figure*}

In Appendix\,\ref{app_sec:D}, we examined the evolution of single-star models accreting beyond the critical rotation limit under the assumption of reduced accreted specific angular momentum. Here, we extend this approach to a toy binary model to test whether our conclusions remain valid in binaries that allows continued accretion after the accretor reaches critical rotation.
As in the single-star Set IV models, we artificially reduce the specific angular momentum of the accreted material by a factor of 10 once the accretor approaches critical rotation. This reduction allows the star to accrete at relatively high rates after reaching critical rotation than in our rotation-limited binary models.
The evolution of this binary system is shown in Fig.\,\ref{app_fig:binary4}.

Compared to the other binary models, the mass-accretion rate in this case declines more slowly after the accretor reaches critical rotation, owing to the reduced specific angular momentum of the accreted material. Nevertheless, the mass transfer remains non-conservative (approximately 10.5\%), since the accreted angular momentum is reduced rather than completely removed.
As in the single-star models in Set IV, a sustained high accretion rate prevents the stellar radius from contracting significantly during accretion. Substantial contraction occurs only once both the mass-transfer and accretion rates drop rapidly in the late phase of interaction. Similar to binary Model II, the accretor begins to spin down before mass transfer ends, as the combination of lower specific angular momentum and declining mass-transfer rate can no longer sustain critical rotation.
After TE is re-established, the accretor settles at a rotation rate of about 52\% of the critical value, following the trend observed in the other models (see Fig.\,\ref{fig:v_k}). This toy model illustrates that our conclusions also hold for binaries in which accretion continues beyond the conventional rotation-limited prescription. We also tested decreasing the specific angular momentum of the accreted material further (by factors of 100 and 1000), allowing for higher maximum accretion rates, but encountered numerical problems once the accretion rate exceeded approximately $6\times10^{-4}\,M_\odot\,\mathrm{yr^{-1}}$. Future studies that treat mass and angular-momentum accretion in a more self-consistent manner are therefore highly desirable.

%NOTE: The accretion timescale is longer than in our single-model IV, therefore the accreted mass is not so large. 

\section{Results for binary model V}\label{app_sec:G}
\setcounter{figure}{0}
\renewcommand\thesection{\Alph{section}}
\renewcommand{\thefigure}{\thesection.\arabic{figure}}
\makeatletter
\renewcommand{\theHfigure}{\thesection.\arabic{figure}} % Update hyperref anchors
\makeatother
We adopt the Tayler–Spruit dynamo mechanism in our models, which provides an efficient way of transporting angular momentum within stars and can lead to nearly rigid rotation. However, its physical viability and efficiency remain debated. Some MHD simulations (e.g., \citealt{2007A&A...474..145Z}) suggest that the Tayler instability alone does not regenerate a sufficiently strong poloidal field to sustain a dynamo loop, or that it saturates at much lower field strengths than predicted by Spruit’s formula. Furthermore, the commonly used assumption that the Tayler–Spruit dynamo operates in a stationary, saturated state may break down during stellar spin-up or spin-down. A more realistic treatment would require time-dependent modelling of magnetic field evolution, as in \cite{2021A&A...646A..19T}, where the Alfvén timescale plays a key role in determining the transport efficiency.

\begin{figure*}
\includegraphics[width=\linewidth]{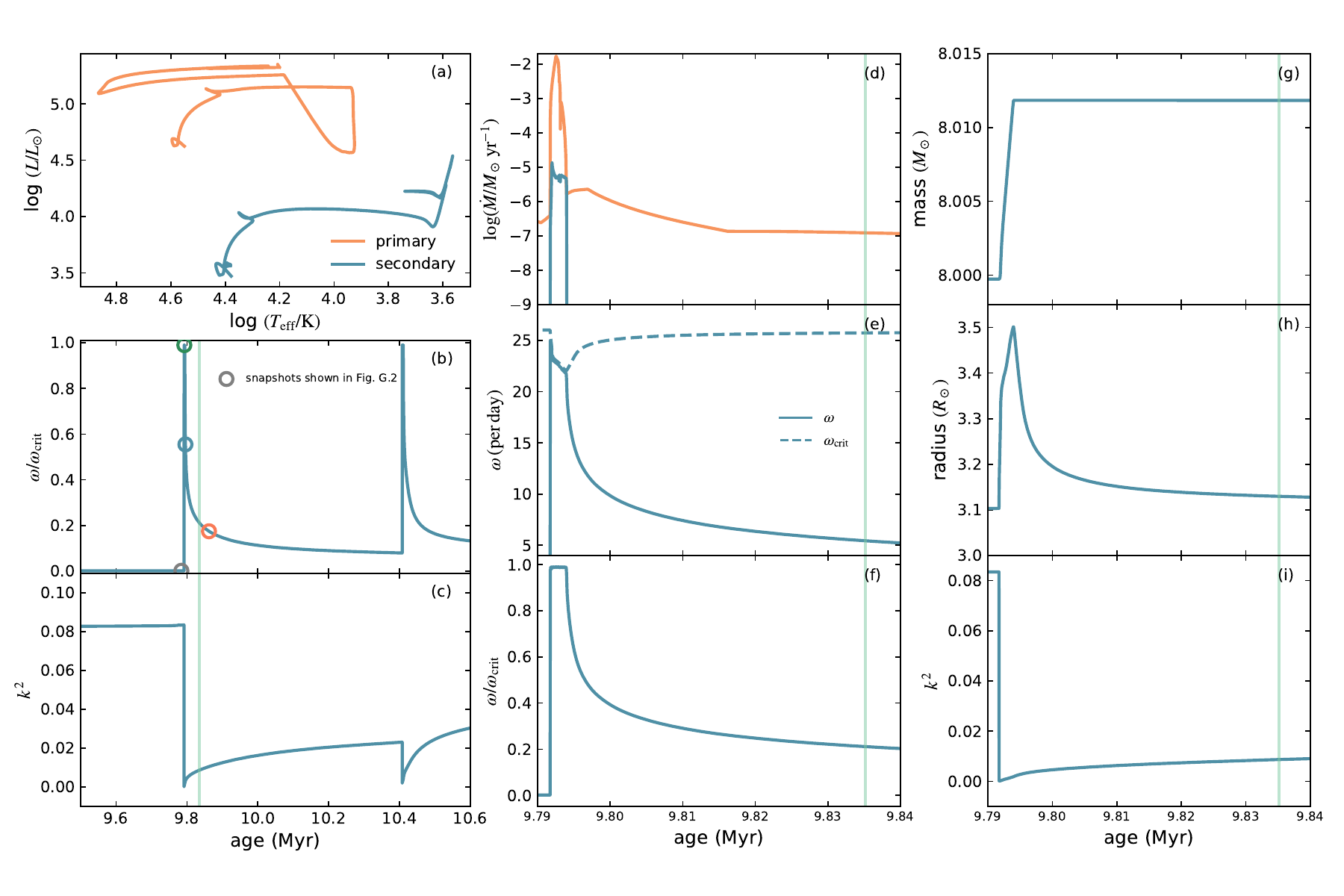}
\caption{Evolution of binary Model V, in which the Tayler–Spruit dynamo is not included, with orange and blue lines showing the primary and secondary stars, respectively.
Panel (a) Evolution in the Hertzsprung-Russell diagram. (b,c) Evolution of $\omega/\omega_\mathrm{crit}$ and $k^2$ over the period encompassing Case B and Case BB mass transfer. (d-i) Stellar properties during Case B mass transfer and the subsequent thermal relaxation phase. Vertical cyan lines in Panels (b) to (i) mark the time when thermal equilibrium is re-established. Open circles in panels (b) indicate the four epochs at which the internal spin profiles are shown in Fig.\,\ref{app_fig:prof_noST}, using the same colors as the corresponding lines in that figure.}
\label{app_fig:ev_noST}
\end{figure*}

\begin{figure}
\includegraphics[width=\linewidth]{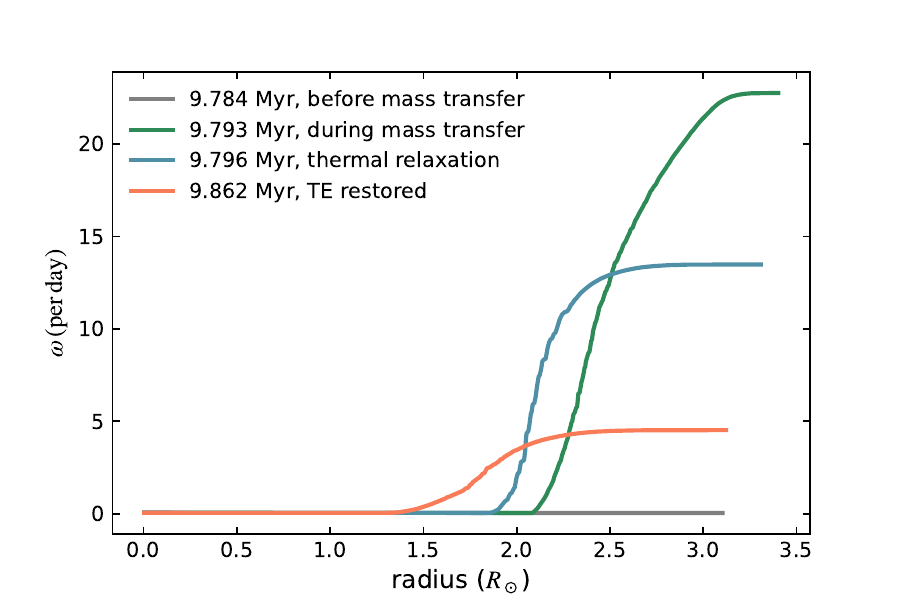}
\caption{Stellar angular-velocity profiles as a function of radius coordinate at four evolutionary stages: stages before mass transfer (grey), during mass transfer (green), during thermal relaxation (blue) and after the restoration of thermal equilibrium (orange). The corresponding epochs are marked with open symbols of the same colors in Fig.\,\ref{app_fig:ev_noST}.}
\label{app_fig:prof_noST}
\end{figure}

A full exploration of realistic angular momentum transport is beyond the scope of this study. Instead, we also examine the opposite extreme case in which the Tayler–Spruit dynamo is absent (binary Model V), resulting in highly inefficient angular momentum transport. The binary evolution and stellar spin profiles at four epochs are shown in Fig.\,\ref{app_fig:ev_noST} and Fig.\,\ref{app_fig:prof_noST}, respectively. During mass transfer, the stellar surface is spun up to critical rotation very quickly. Without the Tayler–Spruit dynamo, angular momentum cannot be efficiently transported into the stellar core, so the surface remains at critical rotation, which prevents further accretion. The accretion efficiency of this model is only approximately 0.1\%. The maximum accretion rate in this model is orders of magnitude lower than in models that include the Tayler–Spruit dynamo. At the end of mass transfer, the star has a critically rotating envelope and a slowly rotating core, and is slightly out of TE. During the subsequent thermal relaxation phase, the star spins down through both thermal readjustment and inward angular momentum transport. Therefore, the star rotates significantly more slowly than models that include the Tayler–Spruit dynamo.

This experiment serves merely as an exploratory test of the extreme case of inefficient angular-momentum transport. Recent asteroseismic studies indicate nearly rigid rotation in late-B and AF type stars \citep{2021RvMP...93a5001A} as well as only small differential rotation in massive stars \citep{2020FrASS...7...70B}, supporting the picture of efficient internal angular-momentum transport. Further investigations in this direction are therefore highly warranted.

%In summary, internal angular momentum transport influences both the mass-accretion phase and the subsequent spin evolution. Without the Tayler–Spruit dynamo, the mass-accretion efficiency is lower than in models that include it, because the envelope reaches critical rotation quickly but cannot transfer angular momentum into the stellar interior. The resulting lower maximum accretion rate also keeps the star closer to TE.  Nevertheless, surface spin-down still occurs due to both thermal relaxation and the slow transport of angular momentum from the surface into the core, eventually producing a slowly rotating star. 

\end{appendix}

\end{document}